\documentclass[aps,reprint,superscriptaddress,longbibliography,nofootinbib,floatfix]{revtex4-2}

\usepackage[T1]{fontenc}
\usepackage[utf8]{inputenc}
\usepackage{amsmath,amssymb,mathtools,bm}
\usepackage{amsthm}
\usepackage{graphicx}
\usepackage{booktabs}
\usepackage{array}
\usepackage{microtype}
\usepackage{xcolor}
\usepackage[colorlinks=true,allcolors=blue!55!black]{hyperref}
\hypersetup{
  pdftitle={Work Statistics of Autonomous Quantum Energy Pumps},
  pdfauthor={Yang Peng}
}

\newcommand{\Tr}{\operatorname{Tr}}
\newcommand{\Var}{\operatorname{Var}}
\newcommand{\Cov}{\operatorname{Cov}}

\newcommand{\dd}{\mathrm{d}}
\newcommand{\ii}{\mathrm{i}}
\newcommand{\e}{\mathrm{e}}
\newcommand{\1}{\mathbb{I}}
\newcommand{\Hil}{\mathcal{H}}

\newcommand{\Hfull}{\hat{\mathbb{H}}}
\newcommand{\Ufull}{\hat{\mathbb{U}}}
\newcommand{\Hs}{\hat{H}}
\newcommand{\Us}{\hat{U}}
\newcommand{\rhoh}{\hat{\rho}}

\newcommand{\om}{\boldsymbol{\omega}}
\newcommand{\bs}{\boldsymbol{s}}

\newcommand{\bchi}{\boldsymbol{\chi}}

\newcommand{\avg}[1]{\left\langle#1\right\rangle}
\newcommand{\comm}[2]{\left[#1,#2\right]}

\newtheorem{theorem}{Theorem}

\begin{document}

\title{Work Statistics of Autonomous Quantum Energy Pumps}

\author{Yang Peng}
\email{yang.peng@csun.edu}
\affiliation{Department of Physics and Astronomy, California State University, Northridge, California 91330, USA}
\affiliation{Institute of Quantum Information and Matter and Department of Physics, California Institute of Technology, Pasadena, California 91125, USA}

\begin{abstract}
We develop a terminal-resolved theory of work statistics for
autonomous quantum energy pumps. By modeling the systems that supply
and receive energy as explicit quantum terminals, the energy exchanged
with each terminal is defined directly from its Hamiltonian change.
When the terminals are modeled by ideal clocks, these full-space observables admit exact
representations on the pump Hilbert space and recover the conventional
phase-derivative currents of periodically and quasiperiodically driven
systems. The framework resolves transported work from energy
accumulated in the pump.  It also incorporates arbitrary
initial pump--terminal correlations and identifies when correlations
can enhance directional energy transfer under uncertain driving
phases. For periodic pumps, we derive finite-cycle work statistics in
Floquet eigenstates, relate their fluctuations to Floquet quantum
geometry, and show that the long-time terminal currents become
mutually compatible. An exactly solvable two-terminal qubit exhibits
noise matching, in which transport becomes sharp through cancellation
of common-mode terminal fluctuations even though the individual
terminal energies remain noisy. Finally, for physical terminals beyond
the ideal-clock limit, we introduce a positive work-variance gap that
quantifies the fluctuations missed by a pump-only description. A
coherent-cavity benchmark shows systematic convergence toward the
ideal-clock regime with increasing occupation while demonstrating that
agreement of the mean current alone does not guarantee accurate work
statistics.
\end{abstract}

\maketitle

\section{Introduction}
\label{sec:introduction}

Coherent energy transfer is becoming a device-level requirement for
quantum technology. Future quantum architectures will combine
components with different transition frequencies, energy scales,
coherence properties, and physical implementations. Connecting these
components requires not only the transfer of quantum information, but
also the controlled supply, conversion, routing, storage, and recovery
of energy with limited noise and dissipation. Quantum transducers
provide a prominent example: by converting excitations between
otherwise incompatible platforms, they enable communication across
heterogeneous quantum networks \cite{Kimble2008,Lauk2020}. Quantum
control, refrigeration, sensing, and computation likewise rely on
energetic resources whose fluctuations become relevant at the
few-quantum level
\cite{Niedenzu2019,Elouard2023,Kurman2026}. A microscopic theory of
such devices must therefore describe not only quantum states and
information, but also the direction, magnitude, and statistics of
energy flow~\cite{Auffeves2022,Campaioli2024}.

Quantum pumping has a long history in particle and charge transport.
Cyclic parameter modulation can generate directed transport without a
static bias, with topology providing quantization and robustness in
suitable adiabatic regimes \cite{Thouless1983,Brouwer1998}. This idea
has more recently been extended to coherent energy conversion between
time-dependent fields. A spin, qubit, cavity, or many-body system
driven at two or more frequencies can absorb quanta from one drive and
emit them into another, thereby acting as a frequency converter or
quantum energy pump. Topological frequency conversion relates this
transfer to a Chern number in a synthetic frequency space
\cite{Martin2017,PengRefael2018,Kolodrubetz2018}. Subsequent work has
addressed nonadiabatic and quasiperiodic pumping, extended systems,
dissipative cavities, and material implementations
\cite{Nathan2019,Crowley2020,Long2021,Qi2021,Long2022,Nathan2022Weyl},
as well as superconducting-qubit Floquet lattices and ensembles of
driven Hamiltonians \cite{MalzSmith2021,Psaroudaki2023}. Recent
developments include finite-speed non-Abelian geometric pumps and
interacting quantum chargers with collectively enhanced mean transfer
rates \cite{Peng2026,Schmid2026}.

Most descriptions of quantum energy pumping begin with a prescribed
time-dependent Hamiltonian. Each external field is specified by its
frequency, phase, and amplitude, and the current assigned to drive
$i$ is obtained by multiplying the phase derivative of the Hamiltonian
by the corresponding frequency~\cite{Martin2017,PengRefael2018,Crowley2020,Qi2021,Long2021,Peng2026,Schmid2026}. This prescription provides a natural
resolution of the average energy flow among classical drives. The
physical sources of those drives, however, are absent from the model.
Their energies and quantum states are not represented, so the driven
description does not directly capture depletion, energetic
backaction, or correlations between the energy sources and the pump.
It also leaves open how the phase-derivative current should be promoted
from a mean-power formula to an observable describing the quantum
statistics of energy exchange.

Moreover, the mean current alone does not fully characterize a
quantum pump. Two protocols with the same mean transferred energy may
have different fluctuations and different correlations between source
loss and receiver gain. The energy retained by the pump can also
fluctuate even when its mean change vanishes after a cycle. These
effects control the precision and reliability of energy conversion but
are invisible to the mean current
\cite{Friis2018,GarciaPintos2020,Bakhshinezhad2024,Mohan2026}. The
situation parallels mesoscopic transport, where current noise and full
counting statistics reveal processes hidden at the level of average
currents \cite{Levitov1996,Landi2024}.

These questions motivate an autonomous formulation in which the
systems supplying and receiving energy are included explicitly as
quantum degrees of freedom. Such bookkeeping is standard in
quantum-battery models, where the charger and battery are treated as
interacting subsystems and charging is characterized by the battery
energy change, supplemented when appropriate by extractable work or
ergotropy \cite{Allahverdyan2004,Andolina2018,Andolina2019,Campaioli2024}. Promoting the
time-dependent drives to quantum energy terminals places driven pumps
and autonomous energy-storage devices within a common Hamiltonian
framework. The work supplied by each terminal is then defined directly
as the decrease of its own energy.

Building on autonomous formulations that relate quantum work to the
energy transferred by explicitly modeled clocks
\cite{Cepollaro2026,Han2025,ZhaoQuan2026}, we develop a
terminal-resolved framework for the joint energy changes of multiple
terminals. The framework distinguishes energy transported through the
pump from energy accumulated in it and incorporates arbitrary initial
pump--terminal correlations for the ideal coordinate-diagonal clock
coupling considered below. In the ideal-clock limit, the autonomous
description recovers the conventional phase-derivative currents of
time-dependent systems. For nonideal physical terminals, the same
energy-bookkeeping framework remains applicable and captures terminal
depletion, dynamical backaction, and pump--terminal correlations
generated during the evolution.

\begin{figure}[t]
 \centering
 \includegraphics[width=\columnwidth]{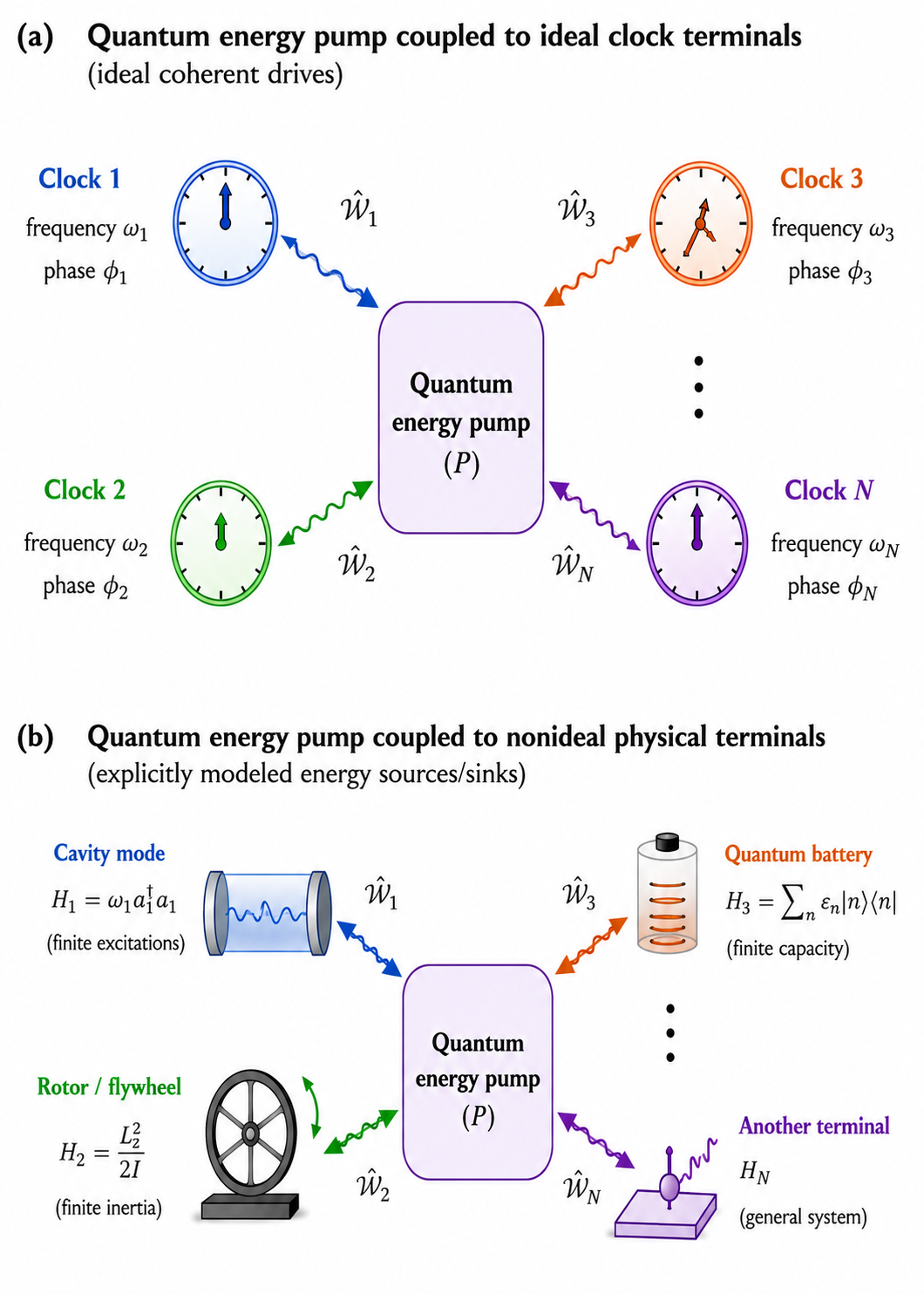}
 \caption{
 Schematic of a quantum energy pump coupled to multiple energy
 terminals. The central pump mediates energy exchange, and the arrows
 denote work supplied by the terminals.
 (a) Ideal clock terminals generate the phases of a periodic or
 quasiperiodic drive.
 (b) Physical terminals beyond the ideal-clock limit may include
 cavity modes, rotors or flywheels, quantum batteries, and other
 energy-bearing quantum systems. Their finite resources and nonrigid
 dynamics can produce depletion and backaction, while pump--terminal
 correlations may modify the transfer statistics.
 }
 \label{fig:pump_terminals}
\end{figure}

Figure~\ref{fig:pump_terminals} illustrates the autonomous setting and
fixes the terminology used throughout. The central system that
mediates energy exchange is the \emph{quantum energy pump}; each
coupled energy-bearing subsystem is an \emph{energy terminal}, or
simply a \emph{terminal}, with its own Hamiltonian. A terminal may
supply or receive energy, and its role may change during the
evolution. We define the work supplied by a terminal as the decrease
of its own energy, so positive terminal work corresponds to energy
delivered to the remainder of the autonomous system.

The paper is organized as follows.
Section~\ref{sec:multiclock} establishes the correspondence between the
autonomous pump--terminal model and the time-dependent description in
the ideal-clock limit. Section~\ref{sec:terminal_work} derives exact
pump-space representations of the terminal-work observables and their
ordered mixed moments, analyzes correlation-assisted pumping under
uncertain clock coordinates, and compares the construction with
photon-resolved Floquet theory (PRFT)
\cite{Engelhardt2024PRR,Engelhardt2024I,Engelhardt2024II,Seibold2025}.
Section~\ref{sec:performance} introduces transport, receiver, and
catalytic performance measures for a two-terminal pump.
Section~\ref{sec:floquet} derives exact repeated-cycle Floquet work
statistics and their quantum-geometric bounds for commensurate drives.
Section~\ref{sec:qubit} illustrates noise matching and
correlation-assisted transfer in an exactly solvable two-clock qubit
model. Section~\ref{sec:general_terminals} extends the framework to
nonideal physical terminals. There, a coherent-cavity realization
quantifies the approach to the ideal-clock limit, and a positive
work-variance gap measures the fluctuations omitted by a pump-only
description. Section~\ref{sec:conclusion} summarizes the results and
outlines future directions.

Throughout, we set $\hbar=1$ and take the work supplied by terminal
$i$ to be positive when that terminal loses energy.

\section{Autonomous pump-terminal dynamics}
\label{sec:multiclock}

Let a quantum energy pump $P$, with Hilbert space $\Hil_P$, interact
with $D$ energy terminals. In this section each terminal is modeled as
an ideal quantum clock. Terminal $i$ has Hilbert space
$\Hil_i=L^2(\mathbb R)$ and generalized coordinate states
$\{|s_i\rangle\}_{s_i\in\mathbb R}$. Its free Hamiltonian generates
rigid translations,
\begin{equation}
 \e^{-\ii\Hs_i t}|s_i\rangle
 =
 |s_i+\omega_i t\rangle,
 \label{eq:clock_translation}
\end{equation}
which is equivalent to $\Hs_i=-\ii\omega_i\partial_{s_i}$ in the
coordinate representation. The parameter $\omega_i$ sets the rate at
which the clock coordinate advances.

For the joint clock system $C$, write
$|\bs\rangle=\bigotimes_{i=1}^D|s_i\rangle$, with
$\bs=(s_1,\ldots,s_D)$ and
$\om=(\omega_1,\ldots,\omega_D)$. The autonomous pump--clock dynamics
is generated by
\begin{equation}
 \Hfull
 =
 \sum_{i=1}^D\Hs_i
 +
 \int_{\mathbb R^D}\dd^D\bs\,
 \Hs_P(\bs)\otimes|\bs\rangle\langle\bs|.
 \label{eq:multi_clock_H}
\end{equation}
Here $\Hs_P(\bs)$ is the pump Hamiltonian conditioned on the clock
coordinates. The coupling is diagonal in the coordinate basis: the
clocks control the pump through their coordinates while translating
rigidly under their free dynamics.

The additive form
\begin{equation}
 \Hs_P(\bs)
 =
 \Hs_0+\sum_{i=1}^D\hat v_i(s_i)
 \label{eq:additive_Hs}
\end{equation}
describes $D$ independently controlled drives. The results below do not
require this restriction; joint dependence on several coordinates
allows nonlinear mixing controlled by multiple terminals.

\begin{theorem}[Multi-terminal autonomization with ideal clocks]
The propagator generated by Eq.~\eqref{eq:multi_clock_H} is
\begin{equation}
 \Ufull(t)
 =
 \int\dd^D\bs\,
 \Us_{\bs}(t)\otimes
 |\bs+\om t\rangle\langle\bs|,
 \label{eq:exact_full_prop}
\end{equation}
where
\begin{equation}
 \Us_{\bs}(t)
 =
 \mathcal T
 \exp\!\left[
 -\ii\int_0^t\dd\tau\,
 \Hs_P(\bs+\om\tau)
 \right].
 \label{eq:conditional_prop}
\end{equation}

For an arbitrary initial pump--clock state $\rhoh_{PC}$, define the
positive operator-valued coordinate density
\begin{equation}
 \hat\sigma_P(\bs)
 =
 \langle\bs|\rhoh_{PC}|\bs\rangle,
 \qquad
 \int\dd^D\bs\,
 \Tr_P\hat\sigma_P(\bs)=1.
 \label{eq:conditional_operator_density}
\end{equation}
The reduced pump state is
\begin{equation}
 \Tr_C\!\left[
 \Ufull(t)\rhoh_{PC}\Ufull^\dagger(t)
 \right]
 =
 \int\dd^D\bs\,
 \Us_{\bs}(t)\hat\sigma_P(\bs)\Us_{\bs}^\dagger(t).
 \label{eq:reduced_correlated}
\end{equation}
For an initial product state
$\rhoh_{PC}=\rhoh_P\otimes\rhoh_C$, one has
$\hat\sigma_P(\bs)=p_C(\bs)\rhoh_P$, where
$p_C(\bs)=\langle\bs|\rhoh_C|\bs\rangle$, and
Eq.~\eqref{eq:reduced_correlated} becomes a statistical mixture of
driven pump evolutions.
\end{theorem}

The proof is given in Appendix~\ref{app:autonomization}.
Equation~\eqref{eq:exact_full_prop} shows that the explicit time
dependence of the driven description arises from the free motion of
clock coordinates under a fixed autonomous Hamiltonian. The terminals
may be mutually correlated and may also be correlated with the pump.
Their coordinate-diagonal blocks then act as coordinate-conditioned
initial pump states. Off-diagonal clock coherences remain present in
the full state but disappear from the reduced pump dynamics because the
coupling and the final partial trace are diagonal in the translated
coordinate basis.

For a sharply localized initial clock configuration
$|\boldsymbol{\phi}\rangle=\bigotimes_{i=1}^D|\phi_i\rangle$, define
\begin{equation}
 \Hs_P^{(\boldsymbol{\phi})}(t)
 \equiv
 \Hs_P(\boldsymbol{\phi}+\om t).
 \label{eq:driven_H}
\end{equation}
The initial coordinates $\boldsymbol{\phi}$ play the role of driving
phases. For the additive coupling in Eq.~\eqref{eq:additive_Hs},
$\Hs_P^{(\boldsymbol{\phi})}(t)=\Hs_0+
\sum_i\hat v_i(\phi_i+\omega_i t)$. Thus ideal clock terminals
reproduce periodic or quasiperiodic driven dynamics within a
time-independent autonomous model.

\section{Terminal-resolved autonomous work}
\label{sec:terminal_work}

\subsection{Exact work observables and currents}

With the sign convention introduced above, the work supplied by
terminal $i$ during the interval $[0,t]$ is the negative change of its
energy:
\begin{equation}
 \hat{\mathcal W}_i(t)
 =
 \Hs_i-\Ufull^\dagger(t)\Hs_i\Ufull(t).
 \label{eq:full_work_def}
\end{equation}
The operator $\hat{\mathcal W}_i(t)$ is a Hermitian observable on the
full pump--terminal Hilbert space. The corresponding
Schr\"odinger-picture energy-current operator is
\begin{equation}
 \hat w_i
 =
 -\ii\comm{\Hfull}{\Hs_i}.
 \label{eq:full_current_def}
\end{equation}
Its Heisenberg evolution integrates to the terminal work:
\begin{equation}
 \hat{\mathcal W}_i(t)
 =
 \int_0^t\dd\tau\,
 \Ufull^\dagger(\tau)\hat w_i\Ufull(\tau).
 \label{eq:full_current}
\end{equation}
For the ideal-clock Hamiltonian in
Eq.~\eqref{eq:multi_clock_H}, the current is diagonal in the clock
coordinates:
\begin{equation}
 \hat w_i
 =
 \int\dd^D\bs\,
 \omega_i\partial_{s_i}\Hs_P(\bs)
 \otimes|\bs\rangle\langle\bs|.
 \label{eq:current_coordinate}
\end{equation}

\begin{theorem}[Exact pump-space representation of terminal work]
For ideal translation clocks, the full terminal-work observable is
diagonal in the initial clock-coordinate basis:
\begin{equation}
 \hat{\mathcal W}_i(t)
 =
 \int\dd^D\bs\,
 \hat W_i^{(\bs)}(t)\otimes|\bs\rangle\langle\bs|.
 \label{eq:block_work}
\end{equation}
The corresponding pump-space block is
\begin{align}
 \hat W_i^{(\bs)}(t)
 &=
 \ii\omega_i
 \Us_{\bs}^\dagger(t)
 \partial_{s_i}\Us_{\bs}(t)
 \label{eq:terminal_work_derivative}\\
 &=
 \int_0^t\dd\tau\,
 \Us_{\bs}^\dagger(\tau)
 \left[
 \omega_i\partial_{s_i}
 \Hs_P(\bs+\om\tau)
 \right]
 \Us_{\bs}(\tau).
 \label{eq:source_work_power}
\end{align}
Consequently, for an arbitrary initial pump--clock state
$\rhoh_{PC}$ and any ordered index string
$(i_1,\ldots,i_n)$,
\begin{align}
 &\avg{
 \hat{\mathcal W}_{i_1}(t)\cdots
 \hat{\mathcal W}_{i_n}(t)
 }_{\rhoh_{PC}}
 \nonumber\\
 &\qquad=
 \int\dd^D\bs\,
 \Tr_P\!\left[
 \hat\sigma_P(\bs)
 \hat W_{i_1}^{(\bs)}(t)\cdots
 \hat W_{i_n}^{(\bs)}(t)
 \right],
 \label{eq:mixed_moment_correlated}
\end{align}
where $\hat\sigma_P(\bs)=\langle\bs|\rhoh_{PC}|\bs\rangle$ is the
operator-valued coordinate density introduced in
Eq.~\eqref{eq:conditional_operator_density}.

For the localized product preparation
$\rhoh_{PC}=\rhoh_P\otimes|\bs_0\rangle\langle\bs_0|$,
Eq.~\eqref{eq:mixed_moment_correlated} reduces to
\begin{align}
 &\Tr\!\left[
 \bigl(
 \rhoh_P\otimes|\bs_0\rangle\langle\bs_0|
 \bigr)
 \hat{\mathcal W}_{i_1}(t)\cdots
 \hat{\mathcal W}_{i_n}(t)
 \right]
 \nonumber\\
 &\qquad=
 \Tr_P\!\left[
 \rhoh_P
 \hat W_{i_1}^{(\bs_0)}(t)\cdots
 \hat W_{i_n}^{(\bs_0)}(t)
 \right].
 \label{eq:mixed_moment_exact}
\end{align}
\end{theorem}

A derivation is given in Appendix~\ref{app:workproof}. For a localized
clock preparation with initial coordinates
$\bs=\boldsymbol{\phi}$,
Eq.~\eqref{eq:terminal_work_derivative} becomes
\begin{equation}
 \hat W_i^{(\boldsymbol{\phi})}(t)
 =
 \ii\omega_i
 \Us_{\boldsymbol{\phi}}^\dagger(t)
 \partial_{\phi_i}
 \Us_{\boldsymbol{\phi}}(t).
 \label{eq:driven_terminal_work}
\end{equation}
This is the terminal-resolved work operator in the corresponding
driven description \cite{Schmid2026}. The autonomous construction therefore reproduces every moment of each
terminal-work observable and every ordered mixed moment of the
terminal works.

Equation~\eqref{eq:mixed_moment_correlated} also shows that these
statistics depend only on the coordinate-diagonal density
$\hat\sigma_P(\bs)$. Coherences between distinct clock coordinates
do not contribute for the ideal coordinate-diagonal coupling.
Pump--clock correlations can nevertheless affect the work statistics
because the conditional pump state may vary with $\bs$.

\subsection{Pump--clock correlations and correlation-assisted pumping}
\label{sec:correlated_clocks}

The operator-valued coordinate density contains both the distribution
of initial clock coordinates and the pump state conditioned on those
coordinates. We write
\begin{equation}
 p(\bs)
 =
 \Tr_P\hat\sigma_P(\bs),
 \qquad
 \tilde\sigma_P(\bs)
 =
 \frac{\hat\sigma_P(\bs)}{p(\bs)},
 \label{eq:conditional_states}
\end{equation}
where $\tilde\sigma_P(\bs)$ is defined on the support of $p(\bs)$.
The normalization of $\rhoh_{PC}$ implies
$\int\dd^D\bs\,p(\bs)=1$ and
$\hat\sigma_P(\bs)=p(\bs)\tilde\sigma_P(\bs)$.

For a real direction
$\bm q=(q_1,\ldots,q_D)$ in terminal-work space, define
\begin{equation}
 \hat{\mathcal W}_{\bm q}(t)
 =
 \sum_{i=1}^Dq_i\hat{\mathcal W}_i(t),
 \qquad
 \hat W_{\bm q}^{(\bs)}(t)
 =
 \sum_{i=1}^Dq_i\hat W_i^{(\bs)}(t).
 \label{eq:directional_work}
\end{equation}
The conditional mean directional work at fixed $\bs$ is
\begin{equation}
 m_{\bm q}(\bs,t)
 =
 \Tr_P\!\left[
 \tilde\sigma_P(\bs)
 \hat W_{\bm q}^{(\bs)}(t)
 \right],
 \label{eq:conditional_directional_work}
\end{equation}
and the total mean is
\begin{equation}
 \avg{\hat{\mathcal W}_{\bm q}(t)}
 =
 \int\dd^D\bs\,
 p(\bs)m_{\bm q}(\bs,t).
 \label{eq:conditional_mean}
\end{equation}
The variance obeys the exact decomposition
\begin{align}
 \Var\!\left[
 \hat{\mathcal W}_{\bm q}(t)
 \right]
 &=
 \int\dd^D\bs\,
 p(\bs)
 \Var_{\tilde\sigma_P(\bs)}
 \!\left[
 \hat W_{\bm q}^{(\bs)}(t)
 \right]
 \nonumber\\
 &\quad+
 \Var_p\!\left[
 m_{\bm q}(\bs,t)
 \right].
 \label{eq:law_total_variance}
\end{align}
The first term is the average conditional quantum variance, while the
second is the classical variance of the conditional mean over
$p(\bs)$.

Pump--clock correlations can increase the maximum attainable mean
directional work. Fix the coordinate distribution $p(\bs)$. For a
product preparation, the same pump state must be used for every
$\bs$. Define
\begin{equation}
 \overline{\hat W}_{\bm q}(t)
 =
 \int\dd^D\bs\,
 p(\bs)\hat W_{\bm q}^{(\bs)}(t).
 \label{eq:average_directional_work}
\end{equation}
The corresponding optimum is
\begin{equation}
 \mu_{\rm prod}^{\max}(t)
 =
 \lambda_{\max}\!\left[
 \overline{\hat W}_{\bm q}(t)
 \right],
 \label{eq:product_optimum}
\end{equation}
where $\lambda_{\max}[\hat A]$ denotes the largest eigenvalue of the
Hermitian operator $\hat A$.

A correlated preparation may instead condition the pump state on the
clock coordinates. Optimizing independently at each $\bs$ gives
\begin{equation}
 \mu_{\rm corr}^{\max}(t)
 =
 \int\dd^D\bs\,
 p(\bs)
 \lambda_{\max}\!\left[
 \hat W_{\bm q}^{(\bs)}(t)
 \right].
 \label{eq:correlated_optimum}
\end{equation}
Convexity of the largest eigenvalue implies
\begin{equation}
 \Delta_{\rm corr}(t)
 \equiv
 \mu_{\rm corr}^{\max}(t)
 -
 \mu_{\rm prod}^{\max}(t)
 \geq0.
 \label{eq:correlation_advantage}
\end{equation}
We call $\Delta_{\rm corr}(t)$ the correlation-assisted gain in mean
directional work. A derivation is given in
Appendix~\ref{app:correlationproof}.

The correlated optimum is attained by choosing
$\tilde\sigma_P(\bs)$ within the maximal-eigenvalue eigenspace of
$\hat W_{\bm q}^{(\bs)}(t)$ at each $\bs$. Classical correlations
between the clock coordinates and the pump state are therefore
sufficient. Because the terminal-work observables are diagonal in the
initial clock-coordinate basis, coherence between distinct clock
coordinates does not alter their statistics.

For the locally optimal conditional states, the fixed-coordinate
quantum variance vanishes, and Eq.~\eqref{eq:law_total_variance}
reduces to
\begin{equation}
 \Var_{\rm corr}^{\rm opt}
 \!\left[
 \hat{\mathcal W}_{\bm q}(t)
 \right]
 =
 \Var_p\!\left[
 \lambda_{\max}\!\left(
 \hat W_{\bm q}^{(\bs)}(t)
 \right)
 \right].
 \label{eq:optimal_correlated_variance}
\end{equation}
Residual fluctuations remain whenever the locally optimal work
eigenvalue varies across $p(\bs)$. A concrete illustration is given
by the two-clock qubit pump in Sec.~\ref{sec:qubit} and
Fig.~\ref{fig:correlation_assisted}.

\subsection{Total work and energy balance}

The sum of the terminal-resolved work operators is the net work
delivered to the pump. Using
$\dd\Hs_P(\bs+\om\tau)/\dd\tau=
\sum_i\omega_i\partial_{s_i}\Hs_P(\bs+\om\tau)$ in
Eq.~\eqref{eq:source_work_power} gives
\begin{equation}
 \sum_{i=1}^D
 \hat W_i^{(\bs)}(t)
 =
 \Us_{\bs}^\dagger(t)
 \Hs_P(\bs+\om t)
 \Us_{\bs}(t)
 -
 \Hs_P(\bs).
 \label{eq:sum_total_work}
\end{equation}
The right-hand side is the Heisenberg change of the pump Hamiltonian
along the trajectory generated by the initial clock coordinates
$\bs$.

The corresponding full-space identity is
\begin{align}
 \sum_{i=1}^D
 \hat{\mathcal W}_i(t)
 &=
 \int\dd^D\bs\,
 \Big[
 \Us_{\bs}^\dagger(t)
 \Hs_P(\bs+\om t)
 \Us_{\bs}(t)
 -
 \Hs_P(\bs)
 \Big]
 \nonumber\\
 &\hspace{4em}\otimes
 |\bs\rangle\langle\bs|.
 \label{eq:full_total_work}
\end{align}
Thus, the sum of the terminal works measures the net energy gained by
the pump, whereas the individual terminal works resolve how energy is
exchanged among the terminals. The net pump-energy change may vanish
even when different terminals supply and receive nonzero amounts of
energy.

\subsection{Work incompatibility and clock-coordinate geometry}

Terminal-work operators associated with different terminals need not
commute. At fixed evolution time $t$, define the Hermitian coordinate
generators
\begin{equation}
 \hat A_i^{(\bs)}(t)
 =
 \ii\Us_{\bs}^\dagger(t)
 \partial_{s_i}\Us_{\bs}(t),
 \qquad
 \hat W_i^{(\bs)}(t)
 =
 \omega_i\hat A_i^{(\bs)}(t).
 \label{eq:Ai_def}
\end{equation}
For a localized clock preparation, the coordinates $s_i$ become the
initial phases $\phi_i$.

The Maurer--Cartan relation gives
\begin{equation}
 \partial_{s_i}\hat A_j^{(\bs)}(t)
 -
 \partial_{s_j}\hat A_i^{(\bs)}(t)
 =
 \ii
 \comm{
 \hat A_i^{(\bs)}(t)
 }{
 \hat A_j^{(\bs)}(t)
 }.
 \label{eq:Maurer_Cartan}
\end{equation}
Consequently,
\begin{align}
 \comm{
 \hat W_i^{(\bs)}(t)
 }{
 \hat W_j^{(\bs)}(t)
 }
 &=
 -\ii\omega_i\omega_j
 \Big[
 \partial_{s_i}\hat A_j^{(\bs)}(t)
 -
 \partial_{s_j}\hat A_i^{(\bs)}(t)
 \Big].
 \label{eq:work_commutator}
\end{align}
Noncommutativity is therefore controlled by how the response to one
clock coordinate changes when another coordinate is varied. Since the
full work observables are diagonal in $\bs$, their commutators are
determined block by block.

When the terminal works do not commute, they admit no common
projective measurement and hence no ordinary sharp joint probability
distribution. Their ordered mixed moments remain well defined through
Eq.~\eqref{eq:mixed_moment_correlated}, and every Hermitian
directional combination in Eq.~\eqref{eq:directional_work} has a
well-defined spectral distribution. These directional observables are
sufficient for the transport, accumulation, and fluctuation measures
used below.

\subsection{Relation to photon-resolved Floquet theory}
\label{sec:relation_prft}

Photon-resolved Floquet theory (PRFT) connects semiclassical Floquet
dynamics with the photon statistics of coherent bosonic driving modes
through counting fields
\cite{Engelhardt2024PRR,Engelhardt2024I,Engelhardt2024II}.
The present construction is complementary: it starts from the exact
energy changes of specified autonomous terminals and applies to both
photonic and nonphotonic terminals.

For an initially localized clock preparation
$\boldsymbol{\phi}$, a symmetric PRFT generating function is
\begin{equation}
 \mathcal G_{\rm PRFT}(\bchi)
 =
 \Tr_P\!\left[
 \rhoh_P
 \Us_{\boldsymbol{\phi}-\boldsymbol{\delta}/2}^{\dagger}(t)
 \Us_{\boldsymbol{\phi}+\boldsymbol{\delta}/2}(t)
 \right],
 \label{eq:symmetric_counting}
\end{equation}
where $\delta_i=\omega_i\chi_i$. By comparison, the pump-space
work operators define
\begin{equation}
 \mathcal G_{\rm WO}(\bchi)
 =
 \Tr_P\!\left[
 \rhoh_P
 \exp\!\left(
 -\ii
 \sum_i
 \chi_i
 \hat W_i^{(\boldsymbol{\phi})}(t)
 \right)
 \right].
 \label{eq:work_operator_characteristic}
\end{equation}
Expanding the symmetric phase-shift expression around
$\bchi=0$ gives
\begin{align}
 \mathcal G_{\rm PRFT}(\bchi)
 &=
 1
 -
 \ii
 \left\langle
 \sum_i
 \chi_i
 \hat W_i^{(\boldsymbol{\phi})}(t)
 \right\rangle
 \nonumber\\
 &\quad
 -
 \frac{1}{2}
 \left\langle
 \left[
 \sum_i
 \chi_i
 \hat W_i^{(\boldsymbol{\phi})}(t)
 \right]^2
 \right\rangle
 +
 O(\|\bchi\|^3).
 \label{eq:counting_agreement}
\end{align}
The two constructions therefore agree through the first moments and
symmetrized second moments. They generally differ at third and higher
orders because the phase-shift construction imposes a particular
ordering of the current operators, whereas
Eq.~\eqref{eq:work_operator_characteristic} generates powers of one
Hermitian directional work observable. A derivation of
Eq.~\eqref{eq:counting_agreement} is given in
Appendix~\ref{app:PRFT}.

PRFT is naturally formulated for the photon statistics of coherent
bosonic modes. The present formulation instead describes the spectral
statistics of the exact terminal-energy changes
$\hat{\mathcal W}_i(t)$. The block identity
Eq.~\eqref{eq:block_work} holds at arbitrary finite times and does not
require periodic evolution or a bosonic realization of the terminals.

\section{Two-clock pump performance}
\label{sec:performance}

We now specialize to two terminals and a localized initial clock
configuration $\boldsymbol{\phi}$. Throughout this section we write
$\hat W_i(t)\equiv\hat W_i^{(\boldsymbol{\phi})}(t)$.
With our sign convention, positive $\hat W_i$ denotes energy supplied
by terminal $i$ to the pump. We designate terminal 1 as the source and
terminal 2 as the receiver. Their mean energy loss and gain are
therefore
$\langle\hat W_1\rangle$ and
$-\langle\hat W_2\rangle$, respectively.

Define the transport and accumulation works by
\begin{align}
 \hat W_{\rm tr}(t)
 &=
 \frac{
 \hat W_1(t)-\hat W_2(t)
 }{2},
 \nonumber\\
 \hat W_{\rm acc}(t)
 &=
 \hat W_1(t)+\hat W_2(t).
 \label{eq:tr_acc_def}
\end{align}
The individual terminal works are
\begin{equation}
 \hat W_1
 =
 \hat W_{\rm tr}
 +
 \frac{1}{2}\hat W_{\rm acc},
 \qquad
 -\hat W_2
 =
 \hat W_{\rm tr}
 -
 \frac{1}{2}\hat W_{\rm acc}.
 \label{eq:terminal_decomp}
\end{equation}
Thus, $\hat W_{\rm tr}$ resolves terminal-to-terminal transport,
whereas $\hat W_{\rm acc}$ is the net work retained by the pump.
Only when $\hat W_{\rm acc}=0$ does the transport work coincide
exactly with both the source loss and the receiver gain.

Equation~\eqref{eq:sum_total_work} gives
\begin{equation}
 \hat W_{\rm acc}(t)
 =
 \Us_{\boldsymbol{\phi}}^\dagger(t)
 \Hs_P^{(\boldsymbol{\phi})}(t)
 \Us_{\boldsymbol{\phi}}(t)
 -
 \Hs_P^{(\boldsymbol{\phi})}(0).
 \label{eq:acc_total}
\end{equation}
For a cyclic Hamiltonian and an initial Floquet eigenstate,
$\langle\hat W_{\rm acc}(nT)\rangle=0$ at stroboscopic times, although
$\langle\hat W_{\rm tr}(nT)\rangle$ may be nonzero.

The mean transport power is
\begin{equation}
 P_{\rm tr}(t)
 =
 \frac{
 \langle\hat W_{\rm tr}(t)\rangle
 }{t},
 \label{eq:transport_power}
\end{equation}
and its relative fluctuation is
\begin{equation}
 \epsilon_{\rm tr}(t)
 =
 \frac{
 \sqrt{
 \Var[
 \hat W_{\rm tr}(t)
 ]
 }
 }{
 \left|
 \langle\hat W_{\rm tr}(t)\rangle
 \right|
 }.
 \label{eq:power_precision}
\end{equation}
The receiver-gain relative fluctuation is
\begin{equation}
 \epsilon_{\rm rec}(t)
 =
 \frac{
 \sqrt{
 \Var[
 \hat W_2(t)
 ]
 }
 }{
 \left|
 \langle\hat W_2(t)\rangle
 \right|
 }.
 \label{eq:receiver_precision}
\end{equation}
Transport and receiver fluctuations need not coincide because the
receiver gain contains an accumulation contribution.

To quantify the residual pump-energy change, define the catalytic
error
\begin{equation}
 \epsilon_{\rm cat}(t)
 =
 \frac{
 \sqrt{
 \langle
 \hat W_{\rm acc}^2(t)
 \rangle
 }
 }{
 \left|
 \langle
 \hat W_{\rm tr}(t)
 \rangle
 \right|
 }.
 \label{eq:catalytic_error}
\end{equation}
A small $\epsilon_{\rm cat}$ means that the root-mean-square
accumulation is small compared with the mean transported work. These
relative quantities are understood when their denominators are
nonzero.

For Hermitian operators $\hat A$ and $\hat B$, define the symmetrized
covariance as
\begin{equation}
 \Cov_{\rm s}(\hat A,\hat B)
 =
 \frac{1}{2}
 \left\langle
 \left\{
 \Delta\hat A,
 \Delta\hat B
 \right\}
 \right\rangle,
 \qquad
 \Delta\hat A
 =
 \hat A-\langle\hat A\rangle.
 \label{eq:sym_covariance}
\end{equation}
The transport and accumulation variances satisfy
\begin{align}
 \Var(\hat W_{\rm tr})
 &=
 \frac{1}{4}
 \left[
 \Var(\hat W_1)
 +
 \Var(\hat W_2)
 \right]
 \nonumber\\
 &\quad
 -
 \frac{1}{2}
 \Cov_{\rm s}
 \left(
 \hat W_1,
 \hat W_2
 \right),
 \label{eq:var_tr_sources}\\
 \Var(\hat W_{\rm acc})
 &=
 \Var(\hat W_1)
 +
 \Var(\hat W_2)
 \nonumber\\
 &\quad
 +
 2
 \Cov_{\rm s}
 \left(
 \hat W_1,
 \hat W_2
 \right),
 \label{eq:var_acc_sources}\\
 \Cov_{\rm s}
 \left(
 \hat W_{\rm tr},
 \hat W_{\rm acc}
 \right)
 &=
 \frac{1}{2}
 \left[
 \Var(\hat W_1)
 -
 \Var(\hat W_2)
 \right].
 \label{eq:cov_tr_acc}
\end{align}
Anticorrelation between the signed terminal works suppresses
accumulation noise while generally enhancing transport noise. Conversely,
sharp transport does not imply a sharp receiver gain when the
accumulation work remains noisy.

Because
$[\hat W_{\rm tr},\hat W_{\rm acc}]=[\hat W_1,\hat W_2]$,
the Schr\"odinger--Robertson uncertainty relation
\cite{Robertson1929,Schrodinger1930} gives
\begin{align}
 &\Var(\hat W_{\rm tr})
 \Var(\hat W_{\rm acc})
 -
 \Cov_{\rm s}^2
 \left(
 \hat W_{\rm tr},
 \hat W_{\rm acc}
 \right)
 \nonumber\\
 &\qquad\geq
 \frac{1}{4}
 \left|
 \left\langle
 \comm{
 \hat W_1
 }{
 \hat W_2
 }
 \right\rangle
 \right|^2.
 \label{eq:SR_pump}
\end{align}
Noncommuting terminal works therefore constrain the simultaneous
precision of transport and accumulation.

\section{Exact Floquet work statistics}
\label{sec:floquet}

We now consider commensurate clock frequencies
$\omega_i=p_i\Omega$, where $p_i$ are integers and
$T=2\pi/\Omega$ is a common period. For a localized initial clock
configuration $\boldsymbol{\phi}$, we again write
$\hat W_i(t)\equiv\hat W_i^{(\boldsymbol{\phi})}(t)$.
The one-period Floquet operator is
\begin{equation}
 \hat F(\boldsymbol{\phi})
 =
 \Us_{\boldsymbol{\phi}}(T).
 \label{eq:Floquet_F}
\end{equation}

\subsection{Time-origin covariance and current conservation}

A common phase shift
$\boldsymbol{\phi}\rightarrow\boldsymbol{\phi}+\om\tau$
corresponds to a change of time origin. Periodicity implies
\begin{equation}
 \hat F(\boldsymbol{\phi}+\om\tau)
 =
 \Us_{\boldsymbol{\phi}}(\tau)
 \hat F(\boldsymbol{\phi})
 \Us_{\boldsymbol{\phi}}^\dagger(\tau).
 \label{eq:time_origin_covariance}
\end{equation}
The Floquet eigenvalues are therefore invariant along the common-time
direction, up to the choice of quasiphase branch. For a smooth,
nondegenerate Floquet band,
\begin{equation}
 \hat F(\boldsymbol{\phi})|\alpha\rangle
 =
 \e^{-\ii\theta_\alpha(\boldsymbol{\phi})}
 |\alpha\rangle,
 \label{eq:Floquet_eigenstate}
\end{equation}
where a smooth local quasiphase branch is understood. It follows that
\begin{equation}
 \sum_{i=1}^D
 \omega_i
 \partial_{\phi_i}
 \theta_\alpha
 =
 0.
 \label{eq:phase_sum_rule}
\end{equation}

Define the cycle-averaged current supplied by terminal $i$ in band
$\alpha$ as
\begin{equation}
 \overline{w}_{i,\alpha}
 =
 \frac{\omega_i}{T}
 \partial_{\phi_i}
 \theta_\alpha.
 \label{eq:band_current}
\end{equation}
Equation~\eqref{eq:phase_sum_rule} then gives
\begin{equation}
 \sum_{i=1}^D
 \overline{w}_{i,\alpha}
 =
 0.
 \label{eq:band_current_conservation}
\end{equation}
Thus, the secular terminal work in a Floquet eigenstate describes
energy redistribution among the terminals without net accumulation in
the pump.

\subsection{Repeated-cycle work statistics}

The one-cycle phase generator associated with terminal $i$ is
\begin{equation}
 \hat A_i(\boldsymbol{\phi})
 =
 \ii
 \hat F^\dagger(\boldsymbol{\phi})
 \partial_{\phi_i}
 \hat F(\boldsymbol{\phi}),
 \label{eq:Floquet_phase_generator}
\end{equation}
so that
\begin{equation}
 \hat W_i(T)
 =
 \omega_i
 \hat A_i(\boldsymbol{\phi}).
 \label{eq:one_cycle_work}
\end{equation}
After $n$ cycles,
\begin{align}
 \hat W_i(nT)
 &=
 \ii\omega_i
 \bigl[
 \hat F^n
 \bigr]^\dagger
 \partial_{\phi_i}
 \hat F^n
 \nonumber\\
 &=
 \omega_i
 \sum_{r=0}^{n-1}
 \hat F^{-r}
 \hat A_i
 \hat F^r.
 \label{eq:ncycle_work_sum}
\end{align}

For a real direction
$\bm q=(q_1,\ldots,q_D)$, define
\begin{equation}
 \hat W_{\bm q}(t)
 =
 \sum_{i=1}^D
 q_i\hat W_i(t)
 \label{eq:Floquet_directional_work}
\end{equation}
and
\begin{equation}
 \partial_{\bm q}
 =
 \sum_{i=1}^D
 q_i\omega_i
 \partial_{\phi_i}.
 \label{eq:directional_derivative}
\end{equation}
The cycle-averaged directional current in band $\alpha$ is
\begin{equation}
 \overline{w}_{\bm q,\alpha}
 =
 \sum_{i=1}^D
 q_i\overline{w}_{i,\alpha}
 =
 \frac{1}{T}
 \partial_{\bm q}
 \theta_\alpha.
 \label{eq:directional_band_current}
\end{equation}

\begin{theorem}[Floquet work statistics]
For a nondegenerate Floquet eigenstate $|\alpha\rangle$, the mean
directional work after $n$ cycles is
\begin{equation}
 \avg{
 \hat W_{\bm q}(nT)
 }_\alpha
 =
 nT\,
 \overline{w}_{\bm q,\alpha}.
 \label{eq:ncycle_mean}
\end{equation}
Its variance is
\begin{align}
 \Var_\alpha\!\left[
 \hat W_{\bm q}(nT)
 \right]
 &=
 4
 \sum_{\beta\neq\alpha}
 \sin^2\!\left[
 \frac{
 n(\theta_\alpha-\theta_\beta)
 }{2}
 \right]
 \nonumber\\
 &\quad\times
 \left|
 \langle\beta|
 \partial_{\bm q}
 \alpha\rangle
 \right|^2.
 \label{eq:ncycle_variance}
\end{align}
\end{theorem}

A proof is given in Appendix~\ref{app:floquetproof}. The directional
Floquet quantum metric is
\begin{equation}
 g_{\bm q\bm q}^{(\alpha)}
 =
 \langle
 \partial_{\bm q}\alpha
 |
 \left(
 \1-|\alpha\rangle\langle\alpha|
 \right)
 |
 \partial_{\bm q}\alpha
 \rangle.
 \label{eq:directional_metric}
\end{equation}
Equation~\eqref{eq:ncycle_variance} immediately gives
\begin{equation}
 \Var_\alpha\!\left[
 \hat W_{\bm q}(nT)
 \right]
 \leq
 4g_{\bm q\bm q}^{(\alpha)}.
 \label{eq:metric_bound}
\end{equation}
The mean work in a Floquet eigenstate grows linearly with the number of
cycles, whereas its variance remains bounded and oscillatory. The
finite-cycle fluctuations are controlled by the quantum geometry of
the Floquet eigenstate in clock-phase space
\cite{Kolodrubetz2017}.

For a finite-dimensional pump with a nondegenerate Floquet spectrum,
Eq.~\eqref{eq:ncycle_work_sum} gives
\begin{equation}
 \lim_{n\rightarrow\infty}
 \left\|
 \frac{
 \hat W_i(nT)
 }{
 nT
 }
 -
 \hat w_i^{(\infty)}
 \right\|
 =
 0,
 \label{eq:asymptotic_current_limit}
\end{equation}
where
\begin{equation}
 \hat w_i^{(\infty)}
 =
 \sum_\alpha
 \overline{w}_{i,\alpha}
 |\alpha\rangle\langle\alpha|.
 \label{eq:asymptotic_current}
\end{equation}
All asymptotic current operators are diagonal in the same Floquet basis
and therefore satisfy
\begin{equation}
 \comm{
 \hat w_i^{(\infty)}
 }{
 \hat w_j^{(\infty)}
 }
 =
 0,
 \qquad
 \sum_{i=1}^D
 \hat w_i^{(\infty)}
 =
 0.
\end{equation}
Thus, although finite-time terminal works may be incompatible, their
long-time cycle-averaged currents possess a common spectral
description.

For a direction $\bm q$, define
\begin{equation}
 \hat w_{\bm q}^{(\infty)}
 =
 \sum_{i=1}^D
 q_i
 \hat w_i^{(\infty)}.
 \label{eq:directional_asymptotic_current}
\end{equation}
For a general initial pump state,
\begin{equation}
 \Var\!\left[
 \hat W_{\bm q}(nT)
 \right]
 =
 (nT)^2
 \Var\!\left[
 \hat w_{\bm q}^{(\infty)}
 \right]
 +
 O(n).
 \label{eq:generic_variance_growth}
\end{equation}
The quadratic term reflects classical uncertainty among Floquet bands
with different cycle-averaged currents. It vanishes in a Floquet
eigenstate, leaving the bounded geometric contribution. For degenerate quasiphases, the asymptotic projection must be taken
onto the corresponding Floquet eigenspaces, and the projected current
operators need not commute within a degenerate subspace.

\subsection{Accumulation fluctuations}

For the two-terminal pump, the accumulation work at stroboscopic times
is
\begin{equation}
 \hat W_{\rm acc}(nT)
 =
 \hat F^{-n}
 \Hs_P(\boldsymbol{\phi})
 \hat F^n
 -
 \Hs_P(\boldsymbol{\phi}).
 \label{eq:acc_stroboscopic}
\end{equation}
Its expectation value vanishes in a Floquet eigenstate:
\begin{equation}
 \avg{
 \hat W_{\rm acc}(nT)
 }_\alpha
 =
 0.
 \label{eq:acc_mean_zero}
\end{equation}
The corresponding variance is
\begin{align}
 \Var_\alpha\!\left[
 \hat W_{\rm acc}(nT)
 \right]
 &=
 4
 \sum_{\beta\neq\alpha}
 \sin^2\!\left[
 \frac{
 n(\theta_\alpha-\theta_\beta)
 }{2}
 \right]
 \nonumber\\
 &\quad\times
 \left|
 \langle\beta|
 \Hs_P(\boldsymbol{\phi})
 |\alpha\rangle
 \right|^2
 \nonumber\\
 &\leq
 4
 (\Delta_\alpha H_P)^2,
 \label{eq:acc_variance_bound}
\end{align}
where
\begin{equation}
 (\Delta_\alpha H_P)^2
 =
 \avg{
 \Hs_P^2(\boldsymbol{\phi})
 }_\alpha
 -
 \avg{
 \Hs_P(\boldsymbol{\phi})
 }_\alpha^2.
 \label{eq:pump_energy_variance}
\end{equation}

For a nonzero transport current
$\overline{w}_{{\rm tr},\alpha}$, the catalytic error satisfies
\begin{equation}
 \epsilon_{\rm cat}(nT)
 \leq
 \frac{
 2\Delta_\alpha H_P
 }{
 nT
 \left|
 \overline{w}_{{\rm tr},\alpha}
 \right|
 }.
 \label{eq:cat_bound}
\end{equation}
A finite-dimensional pump prepared in a Floquet eigenstate therefore
becomes increasingly catalytic in a relative sense: both transport
and accumulation fluctuations remain bounded, whereas the mean
transported work grows linearly with the number of cycles.

\section{Exactly solvable two-clock qubit pump}
\label{sec:qubit}

We illustrate the preceding results with a resonantly driven qubit
coupled to two ideal clock terminals. Both clocks advance with
frequency $\Omega$, and the driven pump Hamiltonian is
\begin{align}
 \Hs_P^{(\boldsymbol{\phi})}(t)
 &=
 \frac{\Omega}{2}\sigma_z
 +\sum_{i=1}^{2}g_i
 \left[
 \cos(\Omega t+\phi_i)\sigma_x
 +\sin(\Omega t+\phi_i)\sigma_y
 \right].
 \label{eq:qubit_H}
\end{align}
Although the clocks have the same frequency, they remain distinct
terminals with separately resolved energy changes.

Introduce the complex transverse amplitude
\begin{equation}
 z
 =
 g_1\e^{\ii\phi_1}
 +
 g_2\e^{\ii\phi_2}
 =
 G\e^{\ii\Phi},
 \label{eq:complex_transverse_amplitude}
\end{equation}
where
\begin{equation}
 G
 =
 \sqrt{
 g_1^2+g_2^2+2g_1g_2\cos\delta
 },
 \qquad
 \delta=\phi_1-\phi_2.
 \label{eq:G_definition}
\end{equation}
For $G>0$, the effective phase is defined by
\begin{equation}
 \e^{\ii\Phi}
 =
 \frac{z}{G},
 \qquad
 \Phi=\arg z
 \quad(\mathrm{mod}\ 2\pi).
 \label{eq:Phi_definition}
\end{equation}

With the rotating-frame transformation
$\hat R(t)=\e^{-\ii\Omega t\sigma_z/2}$, the propagator takes the exact
form
\begin{equation}
 \Us_{\boldsymbol{\phi}}(t)
 =
 \hat R(t)\e^{-\ii\hat h t},
 \label{eq:qubit_propagator}
\end{equation}
where
\begin{align}
 \hat h
 &=
 (\operatorname{Re} z)\sigma_x
 +
 (\operatorname{Im} z)\sigma_y
 \nonumber\\
 &=
 G\left(
 \cos\Phi\,\sigma_x
 +
 \sin\Phi\,\sigma_y
 \right).
 \label{eq:rotating_solution}
\end{align}
For the common period $T=2\pi/\Omega$, one has
$\hat R(T)=-\1$, and therefore
\begin{equation}
 \hat F(\boldsymbol{\phi})
 =
 -\e^{-\ii\hat hT}.
 \label{eq:qubit_Floquet_operator}
\end{equation}

For $G>0$, the eigenstates of $\hat h$ may be chosen as
\begin{equation}
 |s\rangle
 =
 \frac{1}{\sqrt{2}}
 \begin{pmatrix}
 1\\
 s\e^{\ii\Phi}
 \end{pmatrix},
 \qquad
 s=\pm1,
 \label{eq:qubit_band_state}
\end{equation}
with $\hat h|s\rangle=sG|s\rangle$. They are also Floquet
eigenstates:
\begin{equation}
 \hat F(\boldsymbol{\phi})|s\rangle
 =
 \e^{-\ii\theta_s}|s\rangle,
 \qquad
 \theta_s
 =
 \pi+sGT
 \quad(\mathrm{mod}\ 2\pi).
 \label{eq:qubit_quasiphase}
\end{equation}

At $G=0$, the transverse fields cancel, so that
$\hat h=0$ and $\hat F=-\1$. The Floquet spectrum is then degenerate,
and $\Phi$ is undefined. All band-resolved formulas involving
$\Phi$ or derivatives of the Floquet eigenstates are therefore
understood for $G>0$.

For this two-terminal model, the transport and accumulation
derivatives are
\begin{equation}
 \partial_{\rm tr}
 =
 \frac{\Omega}{2}
 \left(
 \partial_{\phi_1}-\partial_{\phi_2}
 \right),
 \qquad
 \partial_{\rm acc}
 =
 \Omega
 \left(
 \partial_{\phi_1}+\partial_{\phi_2}
 \right).
 \label{eq:qubit_directional_derivatives}
\end{equation}
Their action on the effective amplitude and phase is
\begin{align}
 \partial_{\rm tr}G
 &=
 -\Omega
 \frac{g_1g_2\sin\delta}{G},
 &
 \partial_{\rm acc}G
 &=
 0,
 \label{eq:directional_G}\\
 \partial_{\rm tr}\Phi
 &=
 \frac{\Omega(g_1^2-g_2^2)}{2G^2},
 &
 \partial_{\rm acc}\Phi
 &=
 \Omega.
 \label{eq:directional_Phi}
\end{align}

The cycle-averaged terminal currents follow from the quasiphase
derivatives:
\begin{equation}
 \overline{w}_{1,s}
 =
 -s\Omega
 \frac{g_1g_2\sin\delta}{G},
 \qquad
 \overline{w}_{2,s}
 =
 -\overline{w}_{1,s}.
 \label{eq:qubit_current}
\end{equation}
Consequently,
\begin{equation}
 \overline{w}_{{\rm tr},s}
 =
 \overline{w}_{1,s},
 \qquad
 \overline{w}_{{\rm acc},s}
 =
 0.
 \label{eq:qubit_transport_current}
\end{equation}
The two terminals therefore exchange energy without secular
accumulation in the pump.

Equation~\eqref{eq:ncycle_variance} gives the exact finite-cycle
variances
\begin{align}
 \Var_s\!\left[
 \hat W_{\rm tr}(nT)
 \right]
 &=
 \frac{
 \Omega^2(g_1^2-g_2^2)^2
 }{
 4G^4
 }
 \sin^2(nGT),
 \label{eq:qubit_var_tr}\\
 \Var_s\!\left[
 \hat W_{\rm acc}(nT)
 \right]
 &=
 \Omega^2\sin^2(nGT).
 \label{eq:qubit_var_acc}
\end{align}
The corresponding mean transported work is
\begin{equation}
 \avg{
 \hat W_{\rm tr}(nT)
 }_s
 =
 nT\,\overline{w}_{{\rm tr},s}.
 \label{eq:qubit_mean_transport}
\end{equation}
Thus, the mean transported work grows linearly with the number of
cycles, whereas the transport and accumulation variances remain
bounded and oscillatory.

\begin{figure}[t]
 \centering
 \includegraphics[width=0.9\columnwidth]{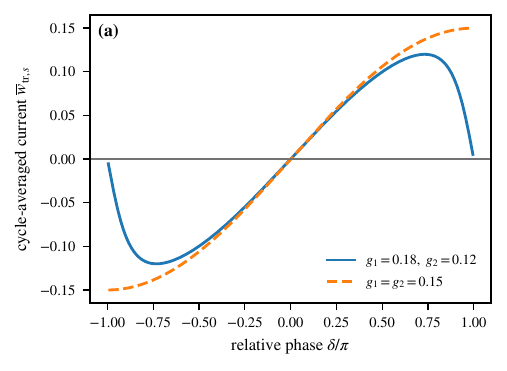}
 \includegraphics[width=0.9\columnwidth]{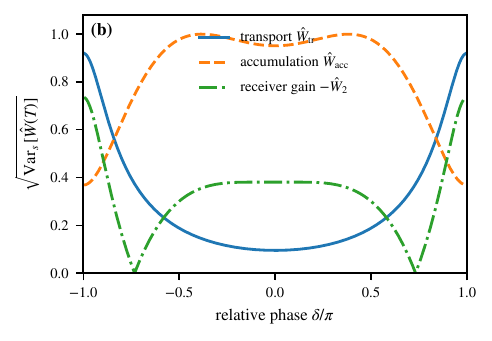}
 \includegraphics[width=0.9\columnwidth]{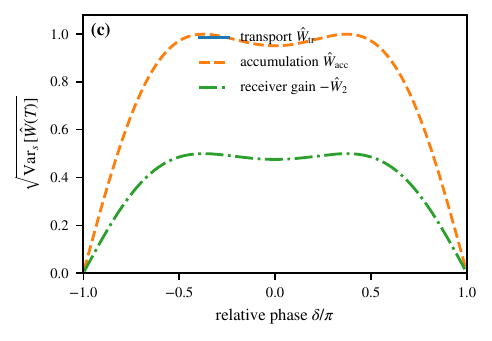}
 \caption{
 Phase dependence of the two-clock qubit pump.
 (a) Cycle-averaged transport current
 $\overline{w}_{{\rm tr},s}$ for unequal couplings
 $(g_1,g_2)=(0.18,0.12)$ and balanced couplings
 $(g_1,g_2)=(0.15,0.15)$.
 (b) One-cycle standard deviations of the transport work
 $\hat W_{\rm tr}$, accumulation work $\hat W_{\rm acc}$, and
 receiver energy gain $-\hat W_2$ for
 $(g_1,g_2)=(0.18,0.12)$.
 (c) The same three standard deviations for
 $(g_1,g_2)=(0.15,0.15)$.
 In all panels, $\Omega=1$, $\phi_1=\delta/2$,
 $\phi_2=-\delta/2$, and $s=-1$.
 The point $\delta=\pi$ in panel (c) is excluded because $G=0$ and
 the Floquet spectrum is degenerate.
 }
 \label{fig:phase_scan}
\end{figure}

Figure~\ref{fig:phase_scan}(a) shows that the relative clock phase
controls the direction and magnitude of the mean energy transfer.
The transport current is odd under
$\delta\rightarrow-\delta$ and changes sign between the two Floquet
bands. For unequal couplings, it vanishes at aligned and anti-aligned
phases and reaches its largest magnitude at intermediate relative
phases. For balanced couplings, the anti-aligned point is instead the
degeneracy $G=0$, where the band-resolved current is not defined.

Panels (b) and (c) of Fig.~\ref{fig:phase_scan} compare the same
three work observables in the unequal- and balanced-coupling regimes.
For unequal couplings, the
transport, accumulation, and receiver-gain observables all generally
fluctuate. The receiver gain contains both transport and accumulation
contributions, as follows from Eq.~\eqref{eq:terminal_decomp}.

For balanced couplings, the transport standard deviation vanishes at
every nondegenerate relative phase, while the accumulation work and
receiver gain remain fluctuating. Phase tuning therefore controls the
mean transfer, whereas coupling balance controls a distinct
noise-matching condition.

Indeed, when $g_1=g_2$ and $G>0$,
\begin{equation}
 \Var_s\!\left[
 \hat W_{\rm tr}(nT)
 \right]
 =
 0
 \label{eq:noise_matching}
\end{equation}
for every cycle number. The transport work is therefore sharp in
either Floquet eigenstate, even though the individual terminal works
need not be sharp. Using Eq.~\eqref{eq:terminal_decomp}, one finds
\begin{align}
 \Var_s\!\left[
 \hat W_1(nT)
 \right]
 &=
 \Var_s\!\left[
 \hat W_2(nT)
 \right]
 \nonumber\\
 &=
 \frac{1}{4}
 \Var_s\!\left[
 \hat W_{\rm acc}(nT)
 \right]
 \nonumber\\
 &=
 \frac{\Omega^2}{4}
 \sin^2(nGT).
 \label{eq:balanced_source_noise}
\end{align}
Equivalently,
\begin{equation}
 \sqrt{
 \Var_s[-\hat W_2(nT)]
 }
 =
 \frac{1}{2}
 \sqrt{
 \Var_s[\hat W_{\rm acc}(nT)]
 }.
 \label{eq:balanced_receiver_noise}
\end{equation}
The two terminal works therefore carry identical common-mode
fluctuations. These fluctuations cancel in the antisymmetric transport
combination while remaining in the accumulation channel. We refer to
this effect as \emph{noise matching}.

\begin{figure}[h]
 \centering
 \includegraphics[width=0.95\columnwidth]{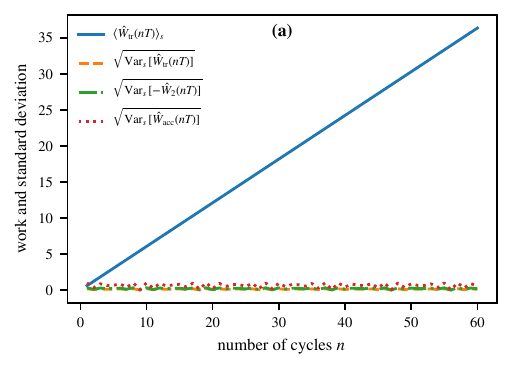}
 \includegraphics[width=0.95\columnwidth]{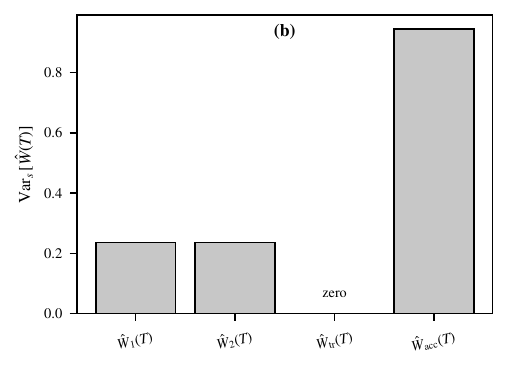}
 \caption{
 Repeated-cycle work statistics and noise matching.
 (a) Mean transported work and standard deviations versus cycle
 number for $\Omega=1$, $(g_1,g_2)=(0.18,0.12)$,
 $(\phi_1,\phi_2)=(0.7,-0.8)$, and $s=-1$.
 (b) One-cycle variances for $\Omega=1$, $g_1=g_2=0.15$,
 $(\phi_1,\phi_2)=(\pi/4,-\pi/4)$, and $s=-1$.
 }
 \label{fig:cycle_results}
\end{figure}

Figure~\ref{fig:cycle_results}(a) illustrates the Floquet scaling
derived in Sec.~\ref{sec:floquet}. The mean transported work
$nT\overline{w}_{{\rm tr},s}$ grows linearly with $n$, whereas the
standard deviations of transport, accumulation, and receiver gain
remain bounded. Their ratios to the mean transported work therefore
decay as $1/n$. The receiver gain retains accumulation fluctuations
through Eq.~\eqref{eq:terminal_decomp} and need not be sharp even
when the transport work is sharp.

Figure~\ref{fig:cycle_results}(b) isolates the balanced-coupling
limit. The transport variance vanishes, while the two terminal
variances are equal to one quarter of the accumulation variance.
Noiseless transport therefore results from cancellation of terminal
fluctuations rather than from the absence of fluctuations in each
terminal.

We finally use the model to illustrate correlation-assisted pumping
under uncertainty in the initial relative phase. We set
$\phi_1=\delta/2$ and $\phi_2=-\delta/2$, and let $\delta$
follow a wrapped Gaussian distribution with mean
$\delta_0$ and width $\Delta\delta$. For a product preparation, the
same initial pump state must be used for every value of $\delta$.
A correlated preparation may instead condition the pump state on the
clock coordinate, as described in
Sec.~\ref{sec:correlated_clocks}.

\begin{figure}[h]
 \centering
 \includegraphics[width=\columnwidth]{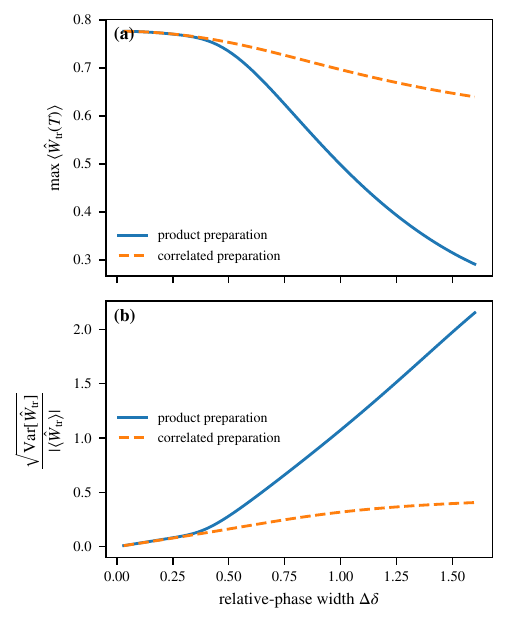}
 \caption{
 Correlation-assisted one-cycle transport under wrapped-Gaussian
 uncertainty in the initial relative phase.
 (a) Maximum attainable mean transport work
 $\mu_{\rm prod}^{\max}(T)$ and
 $\mu_{\rm corr}^{\max}(T)$ for product and correlated pump--clock
 preparations.
 (b) Relative standard deviations of the corresponding optimized
 transport work.
 The parameters are $\Omega=1$, $(g_1,g_2)=(0.18,0.12)$,
 $\delta_0=2$, $\phi_1=\delta/2$, and
 $\phi_2=-\delta/2$.
 The horizontal axis is the phase width $\Delta\delta$.
 }
 \label{fig:correlation_assisted}
\end{figure}

Figure~\ref{fig:correlation_assisted} shows that the product and
correlated optima are nearly equal for a narrow phase distribution.
In this regime, the transport-work operator and its optimal eigenstate
vary little over the occupied clock coordinates. As
$\Delta\delta$ increases, a product preparation must compromise among
operators with different optimal eigenstates. The correlated
preparation can instead select the locally optimal pump state at each
clock coordinate and therefore attains a larger mean transport work
for the displayed parameters.

The optimized fluctuations contain both fixed-coordinate quantum
uncertainty and variation over the phase distribution. For the
locally optimized correlated preparation, the fixed-coordinate
quantum variance vanishes, leaving only the variation of the maximal
work eigenvalue with $\delta$, as expressed by
Eq.~\eqref{eq:optimal_correlated_variance}. The enhancement therefore
requires pump--clock correlations but not coherence between distinct
clock coordinates.

\section{Beyond ideal clocks: coherent-cavity terminals}
\label{sec:general_terminals}

The exact pump-space reduction above relies on ideal translation
clocks: their motion is rigid, their coupling is diagonal in the clock
coordinates, and their terminal-work observables reduce blockwise to
operators on the pump. Physical terminals have finite energy
fluctuations and undergo backaction during operation. Their work
observables remain well defined on the full pump--terminal Hilbert
space, but a pump-only representation is generally approximate. We
first introduce diagnostics for this approximation and then evaluate
them for two coherent cavity terminals coupled to the qubit pump.

\subsection{Exact terminal work and pump-space reduction}

Consider the autonomous Hamiltonian
\begin{equation}
 \Hfull
 =
 \Hs_P+\sum_{i=1}^D\Hs_i+\hat V,
 \label{eq:generic_autonomous_H}
\end{equation}
where $\hat V$ couples the pump to the terminals. The work supplied by
terminal $i$ remains the exact full-space observable defined in
Eq.~\eqref{eq:full_work_def}. For two terminals, we use
\begin{equation}
 \hat{\mathcal W}_{\rm tr}(t)
 =
 \frac{
 \hat{\mathcal W}_1(t)-\hat{\mathcal W}_2(t)
 }{2},
 \qquad
 \hat{\mathcal W}_{\rm acc}(t)
 =
 \hat{\mathcal W}_1(t)+\hat{\mathcal W}_2(t).
 \label{eq:generic_tr_acc}
\end{equation}
Conservation of the total Hamiltonian gives
\begin{equation}
 \hat{\mathcal W}_{\rm acc}(t)
 =
 \Ufull^\dagger(t)
 \left(
 \Hs_P+\hat V
 \right)
 \Ufull(t)
 -
 \left(
 \Hs_P+\hat V
 \right).
 \label{eq:generic_accumulation}
\end{equation}
Thus, for physical terminals, accumulation includes the change of
interaction energy in addition to the change of the bare pump energy.

We next ask how much of the terminal-work statistics can be represented
by an operator acting only on the pump. For an initial product state
$\rhoh_P\otimes\rhoh_C$, where $C$ denotes the joint terminal system,
define the unital completely positive map
\begin{equation}
 \Phi_C(\hat X)
 =
 \Tr_C\!\left[
 \left(
 \1_P\otimes\rhoh_C
 \right)
 \hat X
 \right].
 \label{eq:PhiC}
\end{equation}
For a real direction
$\bm q=(q_1,\ldots,q_D)$ in terminal-work space, let
\begin{equation}
 \hat{\mathcal W}_{\bm q}(t)
 =
 \sum_{i=1}^D
 q_i\hat{\mathcal W}_i(t),
 \qquad
 \hat M_{\bm q}(t)
 =
 \Phi_C\!\left[
 \hat{\mathcal W}_{\bm q}(t)
 \right].
 \label{eq:reduced_first_moment}
\end{equation}
The reduced operator $\hat M_{\bm q}(t)$ reproduces the exact mean
directional work for every initial pump state. Its failure to reproduce
the second moment is measured by the work-variance gap
\begin{equation}
 \hat{\mathcal G}_{\bm q}(t)
 =
 \Phi_C\!\left[
 \hat{\mathcal W}_{\bm q}^2(t)
 \right]
 -
 \hat M_{\bm q}^2(t)
 \geq0,
 \label{eq:directional_gap}
\end{equation}
where the inequality denotes positive semidefiniteness. Positivity
follows from Kadison's inequality \cite{Kadison1952}; see
Appendix~\ref{app:terminalproof}.

The exact variance then separates as
\begin{align}
 \Var_{\rhoh_P\otimes\rhoh_C}
 \!\left[
 \hat{\mathcal W}_{\bm q}(t)
 \right]
 &=
 \Var_{\rhoh_P}
 \!\left[
 \hat M_{\bm q}(t)
 \right]
 \nonumber\\
 &\quad+
 \Tr_P\!\left[
 \rhoh_P\hat{\mathcal G}_{\bm q}(t)
 \right].
 \label{eq:variance_gap_decomp}
\end{align}
The second term is the contribution omitted by the pump-only
operator. For a specified initial pump state, we quantify its relative
importance by
\begin{equation}
 \zeta_{\bm q}(t)
 =
 \frac{
 \Tr_P\!\left[
 \rhoh_P\hat{\mathcal G}_{\bm q}(t)
 \right]
 }{
 \Var_{\rhoh_P\otimes\rhoh_C}
 \!\left[
 \hat{\mathcal W}_{\bm q}(t)
 \right]
 }.
 \label{eq:state_gap_fraction}
\end{equation}
Thus, $\zeta_{\bm q}$ is the fraction of the exact directional-work
variance arising from terminal degrees of freedom that are absent from
the reduced pump operator. It is understood only when the exact
variance is nonzero.

\subsection{Coherent-cavity benchmark}
\label{sec:cavity_benchmark}

We now examine how two coherent cavity terminals approach the
ideal-clock description of the qubit pump in
Sec.~\ref{sec:qubit}. The microscopic Hamiltonian is
\begin{align}
 \Hfull^{\rm cav}
 &={}
 \frac{\Omega}{2}\sigma_z
 +
 \Omega\sum_{i=1}^2
 \hat a_i^\dagger\hat a_i
 \nonumber\\
 &\quad+
 \sum_{i=1}^2
 \lambda_i
 \left(
 \hat a_i\sigma_+
 +
 \hat a_i^\dagger\sigma_-
 \right).
 \label{eq:cavity_model}
\end{align}
Cavity $i$ is initialized in a coherent state with amplitude
$\alpha_i=\sqrt{\bar n_i}\e^{-\ii\phi_i}$, where
$\bar n_i=\avg{\hat a_i^\dagger\hat a_i}$. We scale the microscopic
coupling as  $\lambda_i  =  g_i/\sqrt{\bar n_i}$,
so that
$\lambda_i\alpha_i=g_i\e^{-\ii\phi_i}$ remains fixed as the cavity
occupation increases.

For the numerical benchmark, the cavities have equal occupations,
$\bar n_1=\bar n_2\equiv\bar n$. We take
$\Omega=1$, $(g_1,g_2)=(0.18,0.12)$,
$(\phi_1,\phi_2)=(0.7,-0.8)$, and evolve for one period
$T=2\pi/\Omega$. The matched ideal-clock reference is the two-clock
qubit model of Eq.~\eqref{eq:qubit_H} with the same values of
$\Omega$, $g_i$, and $\phi_i$.

The pump is initialized in the lower Floquet eigenstate of this
reference model, which is $|s\rangle = |s=-1\rangle$ defined in Eq.~\eqref{eq:qubit_band_state}. 
The initial cavity and total states are
\begin{equation}
 \rhoh_C
 =
 |\alpha_1,\alpha_2\rangle
 \langle\alpha_1,\alpha_2|,
 \qquad
 \rhoh_0
 =
 |s\rangle\langle s|
 \otimes
 \rhoh_C.
 \label{eq:cavity_initial_state}
\end{equation}

We use four quantities to compare the cavity dynamics with the
ideal-clock limit. The first is the normalized mean transport
\begin{equation}
 \mathcal R_{\rm tr}(T)
 =
 \frac{
 \Tr_{PC}\!\left[
 \rhoh_0
 \hat{\mathcal W}_{\rm tr}(T)
 \right]
 }{
 \langle s|
 \hat W_{\rm tr}^{\rm id}(T)
 |s\rangle
 }.
 \label{eq:normalized_mean_transport}
\end{equation}
Here
$\hat{\mathcal W}_{\rm tr}$ is constructed from the
exact changes of the two cavity energies, while
$\hat W_{\rm tr}^{\rm id}$ is the transport-work operator of the
matched ideal-clock model. Thus, $\mathcal R_{\rm tr}=1$ means that
the two models give the same mean transported work for the selected
initial pump state.

The second quantity is the terminal-induced fraction of the exact
transport variance,
\begin{equation}
 \zeta_{\rm tr}(T)
 =
 \frac{
 \langle s|
 \hat{\mathcal G}_{\rm tr}(T)
 |s\rangle
 }{
 \Var_{\rhoh_0}\!\left[
 \hat{\mathcal W}_{\rm tr}(T)
 \right]
 }.
 \label{eq:cavity_transport_gap_fraction}
\end{equation}
The positive operator
$\hat{\mathcal G}_{\rm tr}$ is defined in
Eq.~\eqref{eq:directional_gap}. Its expectation gives the part of the
transport variance that is not reproduced by the reduced pump
operator. Therefore, $\zeta_{\rm tr}\to0$ indicates convergence of
the transport variance at the level of the selected pump state.

The remaining two quantities characterize the pump after one period.
The reduced pump state is
\begin{equation}
 \rhoh_P(T)
 =
 \Tr_C\!\left[
 \Ufull^{\rm cav}(T)
 \rhoh_0
 \Ufull^{{\rm cav}\dagger}(T)
 \right],
 \label{eq:reduced_pump_cavity}
\end{equation}
where $\Ufull^{\rm cav}$ is the time-evolution operator defined from the Hamiltonian in Eq.~(\ref{eq:cavity_model}).
We define its return fidelity and linear entropy as
\begin{gather}
 F_P(T)
 =
 \langle s|\rhoh_P(T)|s\rangle,
 \nonumber \\
 S_{L,P}(T)
 =
 1-\Tr_P[\rhoh_P^2(T)].
 \label{eq:cavity_pump_diagnostics}
\end{gather}
The infidelity $1-F_P(T)$ measures departure from the target Floquet
state. The linear entropy measures loss of pump purity
\cite{ManfrediFeix2000}. Because the initial total state is pure and
the evolution is unitary, nonzero $S_{L,P}(T)$ results from
pump--terminal entanglement \cite{Rungta2001}. The ideal-clock limit
is therefore characterized by
$\mathcal R_{\rm tr}\to1$ and
$\zeta_{\rm tr},1-F_P,S_{L,P}\to0$.

To evaluate these quantities efficiently at large occupation, we
introduce the bright and dark modes
\begin{equation}
 \hat b
 =
 \frac{g_1\hat a_1+g_2\hat a_2}{g},
 \qquad
 \hat d
 =
 \frac{-g_2\hat a_1+g_1\hat a_2}{g},
 \label{eq:bright_dark_modes}
\end{equation}
where $g=\sqrt{g_1^2+g_2^2}$.
Only the bright mode couples to the qubit, with coupling
$g/\sqrt{\bar n}$, while the dark mode evolves freely. Its moments
entering the cavity-energy changes are evaluated analytically. The
remaining qubit--bright-mode evolution is obtained from numerically
converged sparse evolution of the truncated Hamiltonian.

The bright-mode Fock space is truncated at
\begin{equation}
 N_b
 =
 \left\lceil
 \bar n_b
 +
 10\sqrt{\bar n_b+1}
 +
 30
 \right\rceil,
 \qquad
 \bar n_b
 =
 |\alpha_b|^2,
 \label{eq:bright_cutoff}
\end{equation}
where
$\alpha_b=(g_1\alpha_1+g_2\alpha_2)/g$. A coherent state has
Poissonian number statistics \cite{Glauber1963}. For the parameters
above, $\bar n_b=1.06530\,\bar n$. At $\bar n=10^3$,
$N_b=1422$, and the omitted initial Poisson weight is
$1.50\times10^{-25}$. Increasing the cutoff leaves the reported
results unchanged within the displayed precision.

\begin{figure}[t]
 \centering
 \includegraphics[width=0.95\columnwidth]
 {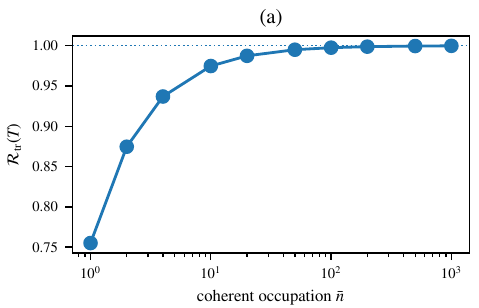}
 \includegraphics[width=0.95\columnwidth]
 {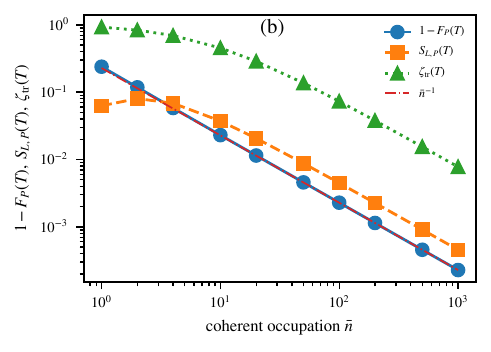}
 \caption{
 Convergence of two coherent cavity terminals toward the ideal-clock
 limit over $1\leq\bar n\leq10^3$.
 (a) Normalized mean transport $\mathcal R_{\rm tr}(T)$.
 (b) Terminal-induced fraction of the exact transport variance
 $\zeta_{\rm tr}(T)$, pump return infidelity $1-F_P(T)$, and pump
 linear entropy $S_{L,P}(T)$.
 The parameters are $\Omega=1$, $(g_1,g_2)=(0.18,0.12)$,
 $(\phi_1,\phi_2)=(0.7,-0.8)$, and the initial pump state is the
 lower ideal-clock Floquet eigenstate.
 The horizontal line in panel (a) marks
 $\mathcal R_{\rm tr}=1$; dash-dotted lines in panel (b) are
 proportional to $\bar n^{-1}$.
 }
 \label{fig:cavity_benchmark}
\end{figure}

Figure~\ref{fig:cavity_benchmark}(a) shows that the physical mean
transport approaches the matched ideal-clock result as the cavity
occupation increases. The deviation
$|\mathcal R_{\rm tr}(T)-1|$ decreases approximately as
$\bar n^{-1}$ over the large-occupation range. In particular,
$\mathcal R_{\rm tr}(T)=0.997472$ at $\bar n=100$ and
$0.999747$ at $\bar n=10^3$.

Figure~\ref{fig:cavity_benchmark}(b) shows that the variance omitted
by a pump-only description, the pump return error, and the generated
pump--terminal entanglement also decrease with occupation. Fits over
$50\leq\bar n\leq10^3$ give approximately inverse-occupation decay:
$\zeta_{\rm tr}\propto\bar n^{-0.9594}$,
$1-F_P\propto\bar n^{-1.0006}$, and
$S_{L,P}\propto\bar n^{-0.9881}$.

At $\bar n=100$, the mean transport is already within approximately
$0.25\%$ of the ideal-clock value, while the eliminated terminal
degrees of freedom still contribute $7.31\%$ of the exact transport
variance. The return infidelity and linear entropy are
$0.228\%$ and $0.447\%$, respectively. At $\bar n=10^3$,
$\zeta_{\rm tr}$ decreases to $0.779\%$, while
$1-F_P=0.0228\%$ and $S_{L,P}=0.0455\%$.

These results show that increasing the coherent occupation suppresses
the disturbance and correlations associated with the physical
terminals and brings both the mean transport and its variance toward
the ideal-clock description. They also illustrate that close
agreement of the mean transport can occur before the omitted
terminal contribution to the variance becomes comparably small.

\subsection{Other nonideal physical terminals}

The same full-space work observables and work-variance-gap construction
apply to the other physical terminals illustrated in
Fig.~\ref{fig:pump_terminals}. Finite battery ladders can approximate
translation only within an operating window limited by spectral
boundaries and dynamical backaction
\cite{Woods2019,Cepollaro2026}. Quantum rotors and mechanical
flywheels provide explicit work-storage terminals, but nonlinear
dispersion can broaden their wave packets and convert directed energy
into passive energy
\cite{Roulet2017,Seah2018,Lindenfels2019}. For finite batteries, the
received energy should likewise be distinguished from its extractable
component, for example through the ergotropy of the final battery
state \cite{AlickiFannes2013}.

For each realization, the exact changes of the terminal Hamiltonians
determine the transported and accumulated work. The positive operator
$\hat{\mathcal G}_{\bm q}$ tests whether the corresponding second
moment can be reproduced by a pump-only work operator, while
$\zeta_{\bm q}$ gives the omitted fraction of the exact variance for a
specified initial pump state. These quantities therefore provide a
common way to assess the validity of a reduced pump-space description
without assuming that the physical terminals realize ideal clocks.

\section{Conclusion and outlook}
\label{sec:conclusion}

We developed a terminal-resolved theory of work statistics for
autonomous quantum energy pumps. The basic observables are the energy
changes $\hat{\mathcal W}_i(t)$ of the individual terminals, rather
than their sum alone. This resolution separates directed transport
from accumulation in the pump and, for physical terminals, in the
interaction sector. For two terminals, the transport work,
accumulation work, source loss, and receiver gain provide distinct
measures of throughput, precision, and catalytic stability.

For ideal translation clocks, the full terminal works admit exact
pump-space blocks. These blocks reproduce every individual work moment
and every ordered mixed moment, and they identify the familiar
phase-derivative current as the exact continuity current of an
explicitly modeled terminal. The autonomous embedding therefore fixes
the terminal label and sign of each current microscopically rather than
introducing them after solving a driven problem. The comparison with
photon-resolved Floquet theory shows agreement through the first and
symmetrized second moments, while higher moments retain different
operator orderings.

The coordinate-block formulation also accommodates arbitrary initial
pump--clock correlations. At fixed clock-coordinate distribution,
conditional pump preparations can attain the pointwise spectral
optimum and outperform every product preparation. The exact law of
total variance separates conditional quantum fluctuations from the
classical variation of the conditional mean. For the diagonal ideal
coupling studied here, however, the work statistics depend only on the
coordinate-diagonal blocks of the initial state. Classical--quantum
correlations are therefore sufficient for the observed advantage, and
the terminal-work statistics do not by themselves witness
pump--terminal entanglement.

For commensurate drives, the Floquet analysis connects work statistics
to quasiphase geometry. In a nondegenerate Floquet eigenstate, the mean
terminal work grows linearly with the number of cycles, while each
directional variance is bounded by the corresponding Floquet quantum
metric. The rescaled terminal works converge in operator norm to a
commuting family of asymptotic current operators. A general initial
state additionally carries a quadratic-in-time variance determined by
its classical uncertainty among Floquet bands. The accumulation work
of a finite-dimensional pump remains bounded, so its root-mean-square
size becomes small relative to the transported work at long times.

The exactly solvable two-clock qubit makes these results explicit. Its
mean current is controlled by the relative terminal phase, whereas its
transport fluctuations are controlled independently by the coupling
asymmetry. At balanced couplings, common-mode terminal fluctuations
cancel from $\hat W_{\rm tr}$ even though the individual terminal works
and receiver gain remain noisy. This noise-matching regime sharpens the
transported work without implying deterministic delivery to the
receiver.

For physical terminals, the full-space energy changes remain exact,
but their complete statistics need not be reproducible by an operator
acting only on the pump. The positive work-variance gap identifies the
part of the second moment omitted by such a reduction, while
$\zeta_{\bm q}$ gives its contribution to the exact variance for a
specified initial pump state. In the coherent-cavity benchmark, the
normalized mean transport approaches the ideal-clock result as the
cavity occupation increases, while the pump return infidelity, the
generated pump--terminal entanglement, and
$\zeta_{\rm tr}$ all decrease. The mean transport can nevertheless be
close to its ideal-clock value while a noticeable fraction of the
exact variance is still missed, showing that agreement at the level of
average energy transfer alone does not validate a pump-only
description of the work statistics.

Several extensions are natural. Finite batteries, rotors, flywheels,
and nonlinear bosonic terminals will introduce spectral boundaries,
dispersion, depletion, and interaction-energy corrections, and their
received energy should be compared with extractable work. Interacting
many-body pumps and terminals may alter both the scaling of the mean
current and the scaling of its fluctuations, generate long-range
terminal correlations, and produce interaction-induced accumulation.
It will be important to distinguish enhanced mean transfer from
improved precision and catalytic stability.

Nondiagonal couplings in the terminal-coordinate basis provide another
promising direction. Such couplings can make the work statistics
sensitive to terminal coherences, generate interference between clock
trajectories, and modify the pump-space moment hierarchy. They may
therefore enable genuinely entanglement-dependent transport advantages
or direct work-statistical witnesses of pump--terminal entanglement.
Further directions include degenerate Floquet manifolds with
non-Abelian asymptotic currents, multi-terminal applications with
$D>2$, and open autonomous devices in which coherent terminal work must be
separated from heat exchanged with uncontrolled environments. These
extensions would broaden terminal-resolved work statistics into a
general device-level theory of throughput, precision, and stability in
quantum energy conversion.

\begin{acknowledgments}
This work is supported by the US National Science Foundation (NSF) Grants  No.\ PHY-2216774 and No.\ DMR-2406524.
\end{acknowledgments}

\appendix

\section{Proof of multi-clock autonomization}
\label{app:autonomization}

We first prove the propagator identity in
Eq.~\eqref{eq:exact_full_prop}. Write
$\Hs_C=\sum_{i=1}^D\Hs_i$ and
\begin{equation*}
 \Hs_{\rm int}
 =
 \int\dd^D\bs\,
 \Hs_P(\bs)\otimes|\bs\rangle\langle\bs|,
\end{equation*}
so that $\Hfull=\Hs_C+\Hs_{\rm int}$. For a short time step
$\Delta t$, the Lie--Trotter formula gives
\begin{align}
 \e^{-\ii\Hfull\Delta t}
 |\psi\rangle|\bs\rangle
 &={}
 \e^{-\ii\Hs_C\Delta t}
 \e^{-\ii\Hs_{\rm int}\Delta t}
 |\psi\rangle|\bs\rangle
 +O(\Delta t^2)
 \nonumber\\
 &={}
 \e^{-\ii\Hs_P(\bs)\Delta t}
 |\psi\rangle
 |\bs+\om\Delta t\rangle
 +O(\Delta t^2).
 \label{eq:Trotter_step}
\end{align}
After $N$ steps, with $t=N\Delta t$,
\begin{align*}
 &\left(
 \e^{-\ii\Hs_C\Delta t}
 \e^{-\ii\Hs_{\rm int}\Delta t}
 \right)^N
 |\psi\rangle|\bs\rangle
 \\
 &\quad={}
 \e^{-\ii\Hs_P(\bs+\om(N-1)\Delta t)\Delta t}
 \cdots
 \e^{-\ii\Hs_P(\bs)\Delta t}
 |\psi\rangle|\bs+\om t\rangle.
\end{align*}
Taking $N\to\infty$ yields
$\Ufull(t)|\psi\rangle|\bs\rangle
 =\Us_{\bs}(t)|\psi\rangle|\bs+\om t\rangle$, where
$\Us_{\bs}(t)$ is the time-ordered propagator in
Eq.~\eqref{eq:conditional_prop}. Linearity in the clock-coordinate
basis then gives Eq.~\eqref{eq:exact_full_prop}.

To derive the reduced pump dynamics, introduce the operator-valued
kernel
$\hat\sigma_P(\bs,\bs')=\langle\bs|\rhoh_{PC}|\bs'\rangle$. Using
Eq.~\eqref{eq:exact_full_prop}, the evolved joint state is
\begin{align*}
 \Ufull(t)\rhoh_{PC}\Ufull^\dagger(t)
 &={}
 \int\dd^D\bs\,\dd^D\bs'\,
 \Us_{\bs}(t)
 \hat\sigma_P(\bs,\bs')
 \Us_{\bs'}^\dagger(t)
 \\
 &\quad\otimes
 |\bs+\om t\rangle
 \langle\bs'+\om t|.
\end{align*}
Tracing over the clocks sets $\bs=\bs'$ and gives
\begin{equation*}
 \Tr_C\!\left[
 \Ufull(t)\rhoh_{PC}\Ufull^\dagger(t)
 \right]
 =
 \int\dd^D\bs\,
 \Us_{\bs}(t)
 \hat\sigma_P(\bs)
 \Us_{\bs}^\dagger(t),
\end{equation*}
where
$\hat\sigma_P(\bs)=\hat\sigma_P(\bs,\bs)
 =\langle\bs|\rhoh_{PC}|\bs\rangle$. This proves
Eq.~\eqref{eq:reduced_correlated}. For an initial product state
$\rhoh_{PC}=\rhoh_P\otimes\rhoh_C$, one has
$\hat\sigma_P(\bs)=p(\bs)\rhoh_P$, with
$p(\bs)=\langle\bs|\rhoh_C|\bs\rangle$, so the reduced dynamics is a
statistical mixture of the conditional pump evolutions.

\section{Proofs for terminal-resolved work}
\label{app:workproof}

\subsection{Terminal-work block identity}

In the clock-coordinate representation, a joint state is a pump-valued
wave function $|\Psi(\bs)\rangle$. Equation~\eqref{eq:exact_full_prop}
acts as
\begin{equation}
 \bigl(\Ufull(t)\Psi\bigr)(\bs)
 =
 \Us_{\bs-\om t}(t)
 |\Psi(\bs-\om t)\rangle.
 \label{eq:prop_wavefunction}
\end{equation}
Using $\Hs_i=-\ii\omega_i\partial_{s_i}$,
\begin{align*}
 \Hs_i\bigl(\Ufull(t)\Psi\bigr)(\bs)
 &={}
 -\ii\omega_i
 \left[
 \partial_{s_i}\Us_{\bs-\om t}(t)
 \right]
 |\Psi(\bs-\om t)\rangle
 \\
 &\quad+
 \Us_{\bs-\om t}(t)
 \Hs_i|\Psi(\bs-\om t)\rangle.
\end{align*}
Applying $\Ufull^\dagger(t)$ and returning to the initial clock
coordinate gives
\begin{align*}
 \Ufull^\dagger(t)\Hs_i\Ufull(t)
 &={}
 \Hs_i
 -
 \int\dd^D\bs\,
 \ii\omega_i
 \Us_{\bs}^\dagger(t)
 \partial_{s_i}\Us_{\bs}(t)
 \\
 &\hspace{7em}\otimes
 |\bs\rangle\langle\bs|.
\end{align*}
Subtracting this expression from $\Hs_i$ proves
Eqs.~\eqref{eq:block_work} and
\eqref{eq:terminal_work_derivative}.

To obtain the integrated-current form, differentiate the conditional
propagator with respect to $s_i$:
\begin{align*}
 \partial_{s_i}\Us_{\bs}(t)
 ={}&
 -\ii\int_0^t\dd\tau\,
 \Us_{\bs}(t,\tau)
 \left[
 \partial_{s_i}\Hs_P(\bs+\om\tau)
 \right]
 \Us_{\bs}(\tau,0),
\end{align*}
where $\Us_{\bs}(t,\tau)$ propagates the pump from $\tau$ to $t$
along the trajectory $\bs+\om\tau$. Multiplying by
$\ii\omega_i\Us_{\bs}^\dagger(t)$ and using the composition law gives
\begin{equation*}
 \hat W_i^{(\bs)}(t)
 =
 \int_0^t\dd\tau\,
 \Us_{\bs}^\dagger(\tau)
 \left[
 \omega_i\partial_{s_i}\Hs_P(\bs+\om\tau)
 \right]
 \Us_{\bs}(\tau),
\end{equation*}
which is Eq.~\eqref{eq:source_work_power}.

Because each full terminal-work operator is diagonal in the initial
clock-coordinate basis, every ordered product is diagonal in the same
basis:
\begin{align*}
 &\hat{\mathcal W}_{i_1}(t)\cdots
 \hat{\mathcal W}_{i_n}(t)
 \\
 &\quad={}
 \int\dd^D\bs\,
 \hat W_{i_1}^{(\bs)}(t)\cdots
 \hat W_{i_n}^{(\bs)}(t)
 \otimes|\bs\rangle\langle\bs|.
\end{align*}
Taking the expectation value in an arbitrary initial state
$\rhoh_{PC}$ selects the diagonal block $\hat\sigma_P(\bs)$ and proves
Eq.~\eqref{eq:mixed_moment_correlated}. Specializing to a localized
clock preparation gives Eq.~\eqref{eq:mixed_moment_exact}.

\subsection{Pump--clock correlations}
\label{app:correlationproof}

Using
$\hat\sigma_P(\bs)=p(\bs)\tilde\sigma_P(\bs)$, the first two
moments of the directional work are
\begin{align*}
 \avg{\hat{\mathcal W}_{\bm q}(t)}
 &={}
 \int\dd^D\bs\,
 p(\bs)m_{\bm q}(\bs,t),
 \\
 \avg{\hat{\mathcal W}_{\bm q}^2(t)}
 &={}
 \int\dd^D\bs\,
 p(\bs)
 \Tr_P\!\left[
 \tilde\sigma_P(\bs)
 \bigl(\hat W_{\bm q}^{(\bs)}(t)\bigr)^2
 \right].
\end{align*}
Adding and subtracting
$\int\dd^D\bs\,p(\bs)m_{\bm q}^2(\bs,t)$ in
$\avg{\hat{\mathcal W}_{\bm q}^2(t)}
 -\avg{\hat{\mathcal W}_{\bm q}(t)}^2$ gives the law of total
variance in Eq.~\eqref{eq:law_total_variance}.

For a product preparation,
$\tilde\sigma_P(\bs)=\rhoh_P$ is independent of $\bs$, and
\begin{align*}
 \avg{\hat{\mathcal W}_{\bm q}(t)}
 &={}
 \Tr_P\!\left[
 \rhoh_P\overline{\hat W}_{\bm q}(t)
 \right],
 \\
 \overline{\hat W}_{\bm q}(t)
 &={}
 \int\dd^D\bs\,
 p(\bs)\hat W_{\bm q}^{(\bs)}(t).
\end{align*}
The variational principle for Hermitian operators then gives
Eq.~\eqref{eq:product_optimum}.

For a correlated preparation, the pointwise bound
\begin{equation*}
 \Tr_P\!\left[
 \tilde\sigma_P(\bs)
 \hat W_{\bm q}^{(\bs)}(t)
 \right]
 \leq
 \lambda_{\max}\!\left[
 \hat W_{\bm q}^{(\bs)}(t)
 \right]
\end{equation*}
implies
\begin{equation*}
 \avg{\hat{\mathcal W}_{\bm q}(t)}
 \leq
 \int\dd^D\bs\,
 p(\bs)
 \lambda_{\max}\!\left[
 \hat W_{\bm q}^{(\bs)}(t)
 \right].
\end{equation*}
The bound is attained by choosing $\tilde\sigma_P(\bs)$ within the
maximal-eigenvalue eigenspace of $\hat W_{\bm q}^{(\bs)}(t)$ for each
$\bs$, proving Eq.~\eqref{eq:correlated_optimum}. Convexity of the
largest eigenvalue gives
\begin{align*}
 \lambda_{\max}\!\left[
 \int\dd^D\bs\,
 p(\bs)\hat W_{\bm q}^{(\bs)}(t)
 \right]
 \leq
 \int\dd^D\bs\,
 p(\bs)
 \lambda_{\max}\!\left[
 \hat W_{\bm q}^{(\bs)}(t)
 \right],
\end{align*}
and hence Eq.~\eqref{eq:correlation_advantage}.

For these locally maximizing conditional states,
$\Var_{\tilde\sigma_P(\bs)}[
 \hat W_{\bm q}^{(\bs)}(t)]=0$. Equation~\eqref{eq:optimal_correlated_variance}
then follows directly from Eq.~\eqref{eq:law_total_variance}.

Finally, consider the formal dephasing operation in the initial
clock-coordinate basis,
\begin{equation*}
 \mathcal D_C(\rhoh_{PC})
 =
 \int\dd^D\bs\,
 \hat\sigma_P(\bs)\otimes|\bs\rangle\langle\bs|.
\end{equation*}
Every ordered product of terminal-work observables is block diagonal
in this basis. Therefore $\rhoh_{PC}$ and
$\mathcal D_C(\rhoh_{PC})$ have identical terminal-work moments, which
proves that these statistics are insensitive to off-diagonal
clock-coordinate coherences.

\subsection{Maurer--Cartan identity}
\label{app:Maurer-Cartan}

At fixed $t$, suppress the coordinate and time labels and define
$\hat B_i=\Us^\dagger\partial_{s_i}\Us=-\ii\hat A_i$.
Commutativity of the coordinate derivatives, together with
$\Us^\dagger\Us=\1$, gives
\begin{equation*}
 \partial_{s_i}\hat B_j
 -\partial_{s_j}\hat B_i
 +\comm{\hat B_i}{\hat B_j}
 =0.
\end{equation*}
Substituting $\hat B_i=-\ii\hat A_i$ yields
\begin{equation*}
 \partial_{s_i}\hat A_j
 -\partial_{s_j}\hat A_i
 =
 \ii\comm{\hat A_i}{\hat A_j},
\end{equation*}
which is Eq.~\eqref{eq:Maurer_Cartan}.

\subsection{Expansion of the PRFT counting amplitude}
\label{app:PRFT}

Define
$\partial_{\boldsymbol\delta}
 =\sum_i\delta_i\partial_{\phi_i}$ and
$\Us_{\pm}=\Us_{\boldsymbol\phi\pm\boldsymbol\delta/2}(t)$.
Expanding around $\boldsymbol\delta=0$ gives
\begin{align*}
 \Us_-^\dagger\Us_+
 &={}
 \1
 +\frac{1}{2}
 \left[
 \Us^\dagger\partial_{\boldsymbol\delta}\Us
 -
 \bigl(\partial_{\boldsymbol\delta}\Us^\dagger\bigr)\Us
 \right]
 \\
 &\quad+
 \frac{1}{8}
 \left[
 \Us^\dagger\partial_{\boldsymbol\delta}^2\Us
 +
 \bigl(\partial_{\boldsymbol\delta}^2\Us^\dagger\bigr)\Us
 \right]
 \\
 &\quad-
 \frac{1}{4}
 \bigl(\partial_{\boldsymbol\delta}\Us^\dagger\bigr)
 \bigl(\partial_{\boldsymbol\delta}\Us\bigr)
 +O(\|\boldsymbol\delta\|^3).
\end{align*}
Using the first and second derivatives of $\Us^\dagger\Us=\1$ reduces
this expression to
\begin{equation*}
 \Us_-^\dagger\Us_+
 =
 \1
 -\ii\hat A_{\boldsymbol\delta}
 -\frac{1}{2}\hat A_{\boldsymbol\delta}^2
 +O(\|\boldsymbol\delta\|^3),
 \qquad
 \hat A_{\boldsymbol\delta}
 =
 \ii\Us^\dagger
 \partial_{\boldsymbol\delta}\Us.
\end{equation*}
Because $\delta_i=\omega_i\chi_i$ and
$\hat W_i^{(\boldsymbol\phi)}(t)
 =\omega_i\hat A_i^{(\boldsymbol\phi)}(t)$, one has
$\hat A_{\boldsymbol\delta}
 =\sum_i\chi_i\hat W_i^{(\boldsymbol\phi)}(t)$. Taking the expectation
value in the initial pump state proves Eq.~\eqref{eq:counting_agreement}.

\section{Floquet derivations}
\label{app:floquetproof}

\subsection{Time-origin covariance}

For a localized initial clock configuration $\boldsymbol\phi$, the
driven pump Hamiltonian satisfies
$\Hs_P^{(\boldsymbol\phi+\om\tau)}(t)
 =\Hs_P^{(\boldsymbol\phi)}(t+\tau)$.
Let $\Us_{\boldsymbol\phi}(t_2,t_1)$ denote the corresponding
propagator from $t_1$ to $t_2$. The Floquet operator for the shifted
initial phases is
$\hat F(\boldsymbol\phi+\om\tau)
 =\Us_{\boldsymbol\phi}(T+\tau,\tau)$.
Periodicity and the composition law give
\begin{align*}
 \hat F(\boldsymbol\phi+\om\tau)
 &={}
 \Us_{\boldsymbol\phi}(T+\tau,\tau)
 \\
 &={}
 \Us_{\boldsymbol\phi}(\tau,0)
 \hat F(\boldsymbol\phi)
 \Us_{\boldsymbol\phi}^\dagger(\tau,0),
\end{align*}
which proves Eq.~\eqref{eq:time_origin_covariance}.

The two Floquet operators are unitarily equivalent and hence have the
same quasiphase spectrum. Differentiating a smooth local quasiphase
branch with respect to $\tau$ gives
$\sum_{i=1}^D\omega_i\partial_{\phi_i}\theta_\alpha=0$, which is
Eq.~\eqref{eq:phase_sum_rule}.

\subsection{Repeated-cycle work statistics}

Recall
$\partial_{\bm q}=\sum_{i=1}^Dq_i\omega_i\partial_{\phi_i}$.
The corresponding one-cycle directional work is
\begin{equation}
 \hat A_{\bm q}
 =
 \ii\hat F^\dagger\partial_{\bm q}\hat F
 =
 \hat W_{\bm q}(T).
 \label{eq:directional_one_cycle_generator}
\end{equation}
Differentiating $\hat F^n$ gives
\begin{align}
 \hat W_{\bm q}(nT)
 &={}
 \ii(\hat F^n)^\dagger
 \partial_{\bm q}\hat F^n
 \nonumber\\
 &={}
 \sum_{r=0}^{n-1}
 \hat F^{-r}\hat A_{\bm q}\hat F^r.
 \label{eq:directional_ncycle_sum}
\end{align}

Let $\hat F|\alpha\rangle
 =\e^{-\ii\theta_\alpha}|\alpha\rangle$.
Differentiating this eigenvalue equation and projecting onto
$|\alpha\rangle$ gives
\begin{equation}
 \langle\alpha|\hat A_{\bm q}|\alpha\rangle
 =
 \partial_{\bm q}\theta_\alpha
 =
 T\overline{w}_{\bm q,\alpha}.
 \label{eq:diagonal_phase_generator}
\end{equation}
For $\beta\neq\alpha$, projection onto $|\beta\rangle$ gives
\begin{align}
 \langle\beta|\hat A_{\bm q}|\alpha\rangle
 &={}
 2\e^{-\ii(\theta_\alpha-\theta_\beta)/2}
 \sin\!\left[
 \frac{\theta_\alpha-\theta_\beta}{2}
 \right]
 \langle\beta|\partial_{\bm q}\alpha\rangle.
 \label{eq:offdiag_phase_generator}
\end{align}

Using Eq.~\eqref{eq:directional_ncycle_sum} and summing the geometric
series yields, for $\beta\neq\alpha$,
\begin{align}
 \langle\beta|\hat W_{\bm q}(nT)|\alpha\rangle
 &={}
 2\e^{-\ii n(\theta_\alpha-\theta_\beta)/2}
 \sin\!\left[
 \frac{n(\theta_\alpha-\theta_\beta)}{2}
 \right]
 \nonumber\\
 &\quad\times
 \langle\beta|\partial_{\bm q}\alpha\rangle.
 \label{eq:offdiag_ncycle}
\end{align}
The diagonal matrix element is
$\langle\alpha|\hat W_{\bm q}(nT)|\alpha\rangle
 =n\partial_{\bm q}\theta_\alpha
 =nT\overline{w}_{\bm q,\alpha}$, proving
Eq.~\eqref{eq:ncycle_mean}. Since $\hat W_{\bm q}(nT)$ is Hermitian,
its variance in $|\alpha\rangle$ is the sum of the squared
off-diagonal matrix elements:
\begin{align*}
 \Var_\alpha[\hat W_{\bm q}(nT)]
 &={}
 4\sum_{\beta\neq\alpha}
 \sin^2\!\left[
 \frac{n(\theta_\alpha-\theta_\beta)}{2}
 \right]
 \\
 &\quad\times
 \left|
 \langle\beta|\partial_{\bm q}\alpha\rangle
 \right|^2.
\end{align*}
This proves Eq.~\eqref{eq:ncycle_variance}.

\subsection{Asymptotic current operators}

For a finite-dimensional Floquet operator with nondegenerate
eigenvalues, define its ergodic projection by
$\mathcal P_F(\hat A)=\sum_\alpha
 |\alpha\rangle\langle\alpha|\hat A
 |\alpha\rangle\langle\alpha|$. The mean ergodic theorem gives
\begin{equation}
 \lim_{n\to\infty}
 \left\|
 \frac{1}{n}
 \sum_{r=0}^{n-1}
 \hat F^{-r}\hat A\hat F^r
 -\mathcal P_F(\hat A)
 \right\|
 =0
 \label{eq:mean_ergodic_floquet}
\end{equation}
for every pump operator $\hat A$.
Applying this result to $\omega_i\hat A_i/T$ and using
$(\omega_i/T)\langle\alpha|\hat A_i|\alpha\rangle
 =\overline{w}_{i,\alpha}$ gives
\begin{equation*}
 \lim_{n\to\infty}
 \left\|
 \frac{\hat W_i(nT)}{nT}
 -\hat w_i^{(\infty)}
 \right\|
 =0,
 \qquad
 \hat w_i^{(\infty)}
 =
 \sum_\alpha
 \overline{w}_{i,\alpha}
 |\alpha\rangle\langle\alpha|.
\end{equation*}
This proves Eqs.~\eqref{eq:asymptotic_current_limit} and
\eqref{eq:asymptotic_current}.

For a general initial pump state, the diagonal part of
$\hat W_{\bm q}(nT)$ grows as
$nT\hat w_{\bm q}^{(\infty)}$, whereas its off-diagonal matrix elements
remain bounded. Therefore
\begin{equation*}
 \Var[\hat W_{\bm q}(nT)]
 =
 (nT)^2\Var[\hat w_{\bm q}^{(\infty)}]
 +O(n),
\end{equation*}
which is Eq.~\eqref{eq:generic_variance_growth}.

For degenerate Floquet eigenvalues, the ergodic projection becomes
$\sum_\lambda\hat P_\lambda\hat A\hat P_\lambda$, where
$\hat P_\lambda$ projects onto the corresponding eigenspace. The
projected current operators need not commute within a degenerate
subspace.

\section{Reduction map and work-variance gap}
\label{app:terminalproof}

The following statements are written for bounded work observables. They
extend to unbounded observables whenever the relevant moments exist and
the corresponding domain conditions are satisfied.

The reduction map $\Phi_C$ in Eq.~\eqref{eq:PhiC} is normal, completely
positive, and unital. Let
$\hat X=\hat{\mathcal W}_{\bm q}(t)$,
$\hat M=\Phi_C(\hat X)$, and
$\hat{\mathcal G}=\Phi_C(\hat X^2)-\hat M^2$.
Kadison's inequality \cite{Kadison1952} gives
\begin{equation*}
 \Phi_C(\hat X^\dagger\hat X)
- 
 \Phi_C(\hat X^\dagger)\Phi_C(\hat X)\geq 0,
\end{equation*}
where "$\geq$" means positive semidefinite.
Because $\hat X$ is Hermitian, this proves
$\hat{\mathcal G}\geq0$ and hence Eq.~\eqref{eq:directional_gap}.

If $\hat{\mathcal G}=0$, the two multiplicative-domain equalities
coincide because $\hat X=\hat X^\dagger$. Choi's multiplicative-domain
theorem \cite{Choi1974} then implies
\begin{equation}
 \Phi_C(\hat X\hat Y)
 =
 \hat M\Phi_C(\hat Y),
 \qquad
 \Phi_C(\hat Y\hat X)
 =
 \Phi_C(\hat Y)\hat M
 \label{eq:MD_relations}
\end{equation}
for every bounded operator $\hat Y$. Repeated application gives
$\Phi_C[p(\hat X)]=p(\hat M)$ for every polynomial $p$. Thus a
vanishing work-variance gap is sufficient for the pump operator
$\hat M$ to reproduce every polynomial moment of $\hat X$. Conversely,
any exact reduced spectral representation preserves the second moment
and therefore requires $\hat{\mathcal G}=0$.

For an initial pump state $\rhoh_P$,
\begin{align*}
 \Var_{\rhoh_P\otimes\rhoh_C}(\hat X)
 &={}
 \Tr_P[\rhoh_P\Phi_C(\hat X^2)]
 -\Tr_P[\rhoh_P\hat M]^2
 \\
 &={}
 \Var_{\rhoh_P}(\hat M)
 +\Tr_P[\rhoh_P\hat{\mathcal G}],
\end{align*}
which proves Eq.~\eqref{eq:variance_gap_decomp}.

Let $\hat\Pi_{\bm q}(\dd w)$ be the spectral measure of
$\hat X=\hat{\mathcal W}_{\bm q}(t)$. The operators
\begin{equation*}
 \hat E_{\bm q}(\dd w)
 =
 \Phi_C[\hat\Pi_{\bm q}(\dd w)]
\end{equation*}
form a positive-operator-valued measure on the pump and reproduce the
full work distribution for every $\rhoh_P$, with the terminal state
$\rhoh_C$ fixed. If $\hat{\mathcal G}=0$, normality and the
multiplicative-domain property imply that $\Phi_C$ acts as a
$*$-homomorphism on the von Neumann algebra generated by $\hat X$.
Consequently, $\hat E_{\bm q}$ is projection valued and is the spectral
measure of $\hat M$. Conversely, if the induced measure is projection
valued, its first moment is $\hat M$ and its second moment is
$\hat M^2$, so $\hat{\mathcal G}=0$. The gap therefore vanishes if and
only if the exact terminal-work distribution is generated by the
spectral measure of a single pump-space operator.

%

\end{document}